\documentclass[vecphys]{svmult}

\def\ltsim{\hbox{\raise 2pt \hbox {$<$} \kern-1.1em \lower 4pt \hbox {$\sim$}}}
\def\ltapprox{\hbox{\raise 2pt \hbox {$<$} \kern-1.1em \lower 5pt \hbox 
{$\approx$}}}
\def\gtsim{\hbox{\raise 2pt \hbox {$>$} \kern-1.1em \lower 4pt \hbox {$\sim$}}}
\def\gtapprox{\hbox{\raise 2pt \hbox {$>$} \kern-1.1em \lower 5pt \hbox 
{$\approx$}}}
\def\arcsec{$^{\prime\prime}$}

\usepackage{makeidx}         % allows index generation
\usepackage{graphicx}        % standard LaTeX graphics tool
\usepackage{multicol}        % used for the two-column index
\usepackage[bottom]{footmisc}% places footnotes at page bottom

\makeindex

\begin{document}

\title*{Galaxy Clusters in the Radio: Relativistic Plasma 
 and ICM/Radio Galaxy Interaction Processes}
%\title*{Galaxy clusters in radio: non-thermal phenomena
%in the ICM and ICM/galaxy interaction}
%\title*{Galaxy clusters in the radio domain: \\non-thermal phenomena
%in the ICM and in galaxies}
\titlerunning{Galaxy clusters in radio}
\author{Luigina Feretti$^1$ and Gabriele Giovannini$^{1,2}$}
\authorrunning{Luigina Feretti and Gabriele Giovannini} 
\institute{$^1$Istituto di Radioastronomia INAF, Via P. Gobetti 101, 40129 Bologna (Italy)\\
\texttt{lferetti@ira.inaf.it}\\
$^2$Dipartimento di  Astronomia, Universit\'a di Bologna, Via Ranzani 1, 40127 Bologna (Italy)\\
\texttt{gabriele.giovannini@unibo.it}}

\maketitle

\section {Introduction}
\label{s:intro}

Studies at radio wavelengths allow the investigation of important
components of clusters of galaxies. 
The most spectacular aspect of cluster radio
emission is represented by the large-scale diffuse radio sources,
which cannot be obviously associated with any individual galaxy.
These sources indicate the existence of relativistic particles and
magnetic fields in the cluster volume, thus the presence of non-thermal
processes in the hot intracluster medium (ICM).  The
knowledge of the properties of these sources has increased
significantly in recent years, due to higher sensitivity radio images and to
the development of theoretical models.  The importance of these
sources is that they are large scale features, which are related to
other cluster properties in the optical and X-ray domain, and are thus
directly connected to the cluster history and evolution.

The radio emission in clusters  can also originate from
individual galaxies, which have been imaged over the last decades with
sensitive radio telescopes. The emission from radio galaxies
often extends well beyond their optical boundaries, out to hundreds of
kiloparsec, and hence it is expected that the ICM would affect their
structure. This interaction is indeed observed in extreme examples:
the existence of radio galaxies showing distorted structures (tailed
radio sources), and radio sources filling X-ray cavities at the centre
of cooling core clusters.  Finally, 
the cluster environment may play a role in the statistical radio
properties of galaxies, i.e. their probability of forming radio
sources.

The organization of this paper is as follows:
The basic formulae used to derive the age
of synchrotron sources and the equipartition parameters are presented 
in \S~\ref{s:syncgen}, while the observational properties of 
diffuse radio sources are presented in \S~\ref{s:diff}. Then in
\S~\ref{s:partic} we give a general outline of the models of the
relativistic particle origin and re-acceleration; while 
the current results on cluster magnetic fields are described in
\S~\ref{s:bfield}. Finally, 
\S~\ref{s:rg} reports the properties of cluster radio emitting galaxies.

The intrinsic parameters quoted in this paper are computed for a 
$\Lambda$CDM cosmology with $H_0$ = 70 km s$^{-1}$Mpc$^{-1}$,
$\Omega_m$=0.3, and $\Omega_{\Lambda}$=0.7.

\section{Basic formulas from the synchrotron theory}
\label{s:syncgen}

\subsection{Synchrotron radiation}
\label{s:sync1}

The synchrotron emission is produced by the spiralling motion of
relativistic electrons in a magnetic field.  
An electron with energy E=$\gamma m_e c^2$ (where $\gamma$ is the
Lorentz factor) in a magnetic field $\vec{B}$, experiences a
$\vec{v}\times\vec{B}$ force that causes it to follow a helical path
along the field lines, emitting radiation into a cone of half-angle
$\simeq~\gamma^{-1}$ about its instantaneous velocity.  To the
observer, the radiation is essentially a continuum with a fairly
peaked spectrum concentrated near the  frequency

\begin{equation}
\nu_{syn}=\frac{3e}{4\pi m_e^3c^5}(B\sin\theta)\E^2,
\label{sync}
\end{equation}

\noindent
where $\theta$ is the pitch angle between the electron velocity 
and the magnetic field direction.
The synchrotron power emitted by a relativistic electron is 

\begin{equation}
-\frac{dE}{dt}=\frac{2e^4}{3m_e^4c^7}(B\sin\theta)^2E^2.
\label{loss}
\end{equation} 

\noindent 
In c.g.s  units:

\begin{equation}
\nu_{syn}\simeq 6.27 \times 10^{18} (B\sin\theta)E^2
\label{cgs}
\end{equation}

\noindent
$~~~~~~~~~~~~~~~~~~~~~~~~~~~~~~~~~~~~~~\simeq 4.2 \times 10^6 (B\sin\theta) \gamma^2$,

%\newpage

\begin{equation}
-\frac{dE}{dt}\simeq 2.37 \times10^{-3}(B\sin\theta)^2 E^2
\label{cgs2}
\end{equation} 
$~~~~~~~~~~~~~~~~~~~~~~~~~~~~~~~~~~~~~\simeq1.6\times10^{-15}(B\sin\theta)^2\gamma^2$.\\

\medskip\noindent
>From eqn. \ref{cgs}, it is easily derived that electrons 
of $\gamma \simeq 10^3-10^4$ in magnetic fields of $B\simeq 1\mu$ G 
radiate in the radio domain.

The case of astrophysical interest is that of a homogeneous and 
isotropic population of electrons with a
power-law energy distribution, i.e., with the particle density between
$E$ and $E$+d$E$ given by:
\begin{equation}
N(E)dE=N_0E^{-\delta}dE.
\label{powerlaw}
\end{equation}
\noindent
To obtain the total monochromatic  emissivity $J(\nu)$, 
one must integrate over the contributions of all electrons. 
In regions which are optically thin to their own radiation (i.e. without
any internal absorption), the total intensity spectrum
varies as \cite{Blu70}:
\begin{equation}
J(\nu)\propto N_0 (B\sin \theta)^{1+\alpha} \nu^{-\alpha},
\label{spectralind}
\end{equation}
\noindent
therefore it follows a power-law with spectral index related to
the index of the electron energy distribution 
$\alpha=(\delta-1)/2$.

\subsection{Time evolution of the synchrotron spectrum}
\label{s:syncage}

By integrating the expression of the electron energy loss 
(eqn. \ref{loss}) it is found 
that the particle energy decreases with time, as:

\begin{equation}
E= \frac{E_0}{1+b (B\sin\theta)^2 E_0 t},
\end{equation} 

\noindent
where $E_0$ is the initial energy at $t=0$, and 
$b = 2e^4/(3m_e^4c^7)$ = 
2.37 $\times$ 10$^{-3}$ c.g.s units (see eqn. \ref{cgs2}).
Therefore, the particle energy halves after a time $t^{*}$ = 
$[b(B\sin\theta)^2E_0]^{-1}$. 
This is a characteristic time which  can be identified as 
the particle lifetime. Similarly, we can define a characteristic energy
$E^{*}$ = $[b(B\sin\theta)^2t]^{-1}$, such that a particle with energy 
$E_0 > E^{*}$ will lose most of its energy in a time $t^*$.

\begin{figure}[t]
\centering
\includegraphics[height=3.7cm]{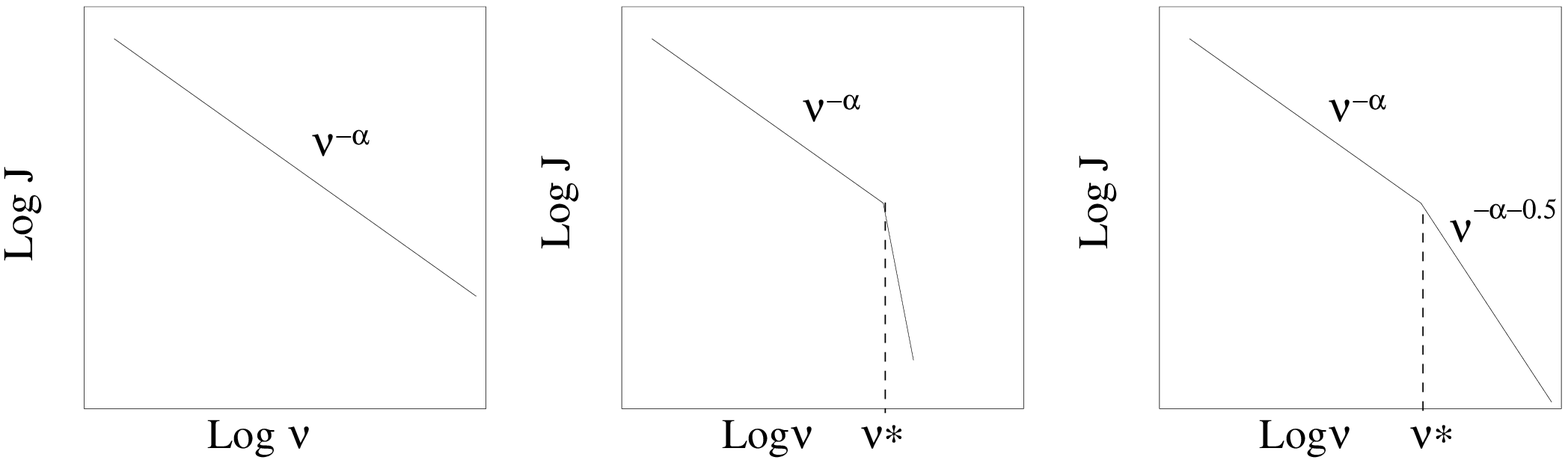}
\caption{
Sketch of synchrotron spectra. The left panel shows a standard
spectrum, the central panel shows an aged spectrum produced
in a source with a single event of particle production, the right panel
shows an aged spectrum with particle injection.
The critical frequency $\nu^*$ is related to the particle lifetime.
}
\label{fig1}       % Give a unique label
\end{figure}

In an ensemble of particles, the energy losses of each particle affect
the overall particle energy distribution, and consequently the resulting
synchrotron spectrum  undergoes a modification. Indeed, after a time
$t^*$ the particles with $E > E^*$ will lose most of their
energy. This produces a critical frequency $\nu^*$ in the radio
spectrum, such that for $\nu <$ $\nu^*$ the spectrum is unchanged,
whereas for $\nu > $ $\nu^*$ the spectrum steepens.  If particles were
produced in a single event with power law energy distribution,
$N(E,0)dE = N_0 E^{-\delta}dE$, the radio spectrum would fall rapidly to
zero for $\nu > $ $\nu^*$.  In the case that new particles were
injected in the source, the spectrum beyond $\nu^*$ steepens by 0.5.
These various cases are illustrated in Fig. \ref{fig1}. Any radio spectrum
showing a cutoff is evidence of ageing of the radio emitting
particles. In addition, any spectrum showing no cutoff but having a
steep spectral index is also indicative of ageing, since it naturally
refers to a range of frequencies higher than the critical frequency.
For a rigorous treatment of the evolution of synchrotron spectra
we refer to \cite{Kar62} and \cite{Van69}.

>From the critical frequency $\nu^*$, it is possible
to derive the radiating electron lifetime, which represents
the time since the particle production (or the time since the last 
injection event, depending on the shape of spectral steepening).
Since the synchrotron emission depends on $\sin\theta$ (eqn. \ref{sync}), 
one has to take into account the distribution of
electron pitch angles.
Moreover, for a correct evaluation, also the electron energy
losses, due to the inverse Compton process, must be considered.

The electron lifetime (in Myr), assuming an {\it anisotropic}
pitch angle distribution is given by:

\begin{equation}
t^*= 1060\frac{B^{0.5}}{B^2 + \frac{2}{3} B^2_{\rm CMB}}\left[ (1+z)\nu^* 
\right]^{-0.5},
\label{anis}
\end{equation} 

\noindent
where the magnetic field B is in $\mu$G, the frequency $\nu$ is
in GHz and B$_{\rm CMB} (= 3.25 \;(1+z)^2$ $\mu$G) is the equivalent
magnetic field of the Cosmic Microwave Background.
If the distribution of electron pitch angles is {\it isotropic}, 
the above formula becomes:

\begin{equation}
t^*= 1590\frac{B^{0.5}}{B^2 +  B^2_{\rm CMB}}\left[ (1+z)\nu^* \right]^{-0.5}.
\label{isot}
\end{equation} 

\noindent
A derivation of the expressions in eqs. \ref{anis} and \ref{isot} 
can be found in \cite{Sle01}.

\subsection{Energy content and equipartition magnetic fields}
\label{s:equip}

The total
energy of a synchrotron source is due to the energy in relativistic
particles ($U_{el}$ in electrons and $U_{pr}$ in protons) plus the
energy in magnetic fields ($U_B$):
\begin{equation}
U_{tot}=U_{el}+U_{pr}+U_B.
\end{equation} 

\noindent
The magnetic field energy contained in the source volume $V$ is given by
\begin{equation} 
U_B = \frac{B^2}{8\pi}\Phi V,
\label{enb}
\end{equation}
\noindent
where $\Phi$ is the fraction of the source volume occupied by the magnetic
field (filling factor).
The electron total energy in the range $E_1$ - $E_2$,
\begin{equation}
 U_{el} =V\times \int_{E_{1}}^{E_{2}} N(E)E \, dE= VN_{0} 
\int_{E_{1}}^{E_{2}} E^{-\delta +1} \,dE 
\label{rif1}
\end{equation}
\noindent
can be expressed as a function of the synchrotron luminosity, $L_{syn}$,
observed between two frequencies $\nu_1$ and $\nu_2$, i.e.,

\begin{equation}
U_{el} = L_{syn} (B\sin\theta)^{-\frac{3}{2}} f(\delta,\nu_1,\nu_2),
\end{equation}

\noindent
where $f(\delta,\nu_1,\nu_2)$ is a function of the index of the electron
energy distribution and of the
 observing frequencies (see \cite{Pac70} for a
rigorous derivation).
The energy contained in the heavy particles, $U_{pr}$, can be related
to $U_{el}$ assuming: 
\begin{equation}
U_{pr} =kU_{el}.
\label{kpar}
\end{equation}
\noindent
Finally, taking $\sin\theta$=1,  the total energy is:
\begin{equation} 
U_{tot} = (1 + k) L_{syn} B^{-\frac{3}{2}} f(\delta,\nu_1,\nu_2) + \frac{B^2}{8\pi}\Phi V.
\label{etot}
\end{equation}

\begin{figure}[t]
\centering
\includegraphics[height=6cm]{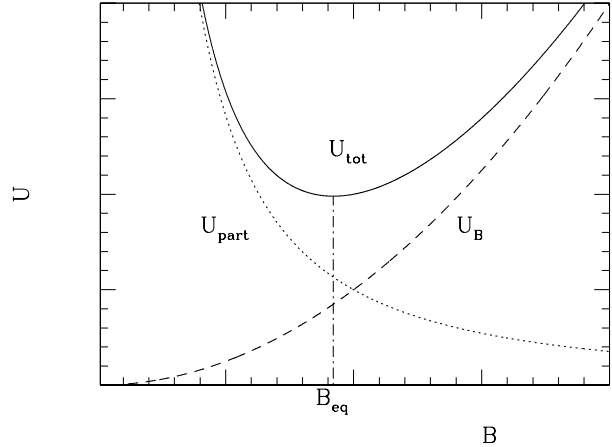}
\caption{
Trend of the energy content in a radio source (in arbitrary units): the
energy in magnetic fields is $U_B \propto$ B$^2$, the energy in
relativistic particles is $U_{part}=U_{el}+U_{pr} \propto$ B$^{-3/2}$.
The total energy content $U_{tot}$ is minimum when the contributions
of magnetic fields and relativistic particles are approximately equal
(equipartition condition). The corresponding magnetic field is
commonly referred to as equipartition value B$_{eq}$.
}
\label{fig2}       % Give a unique label
\end{figure}

\noindent
The trend of the radio source energy content is shown in Fig. \ref{fig2}.
The condition of minimum
energy, $U_{min}$, computed by equating to zero the first derivative
of the expression of $U_{tot}$ (eqn. \ref{etot}), 
is obtained when the contributions of the magnetic field and
the relativistic particles are approximately equal:
\begin{equation} 
U_{B} =\frac{3}{4}(1+k)U_{el}.
\end{equation}
\noindent
For this reason the minimum energy is known also as equipartition
value. 

The total minimum energy density $u_{\rm min}=U_{\rm min}/V \Phi$, 
assuming same volume in particles and magnetic field
($\Phi$=1), and applying the K-correction, can be expressed in terms
of observable parameters, as:

\begin{equation}
u_{\rm min} = 1.23 \times 10^{-12} (1+k)^{\frac{4}{7}}  
(\nu_{0})^{\frac{4\alpha}{7}} (1+z)^{\frac{(12+4\alpha)}{7}} I_0 ^{\frac{4}{7}} d^{\frac{4}{7}},
\label{intnu}
\end{equation}

\noindent
where $I_0$ is the source brightness which is directly observed at the
frequency $\nu_0$, $d$ is the source depth along the line of sight,
$z$ is the source redshift and $\alpha$ is the spectral index of the
radio emission.  The energy density is in erg cm$^{-3}$,
$\nu_{0}$ in MHz, $I_0$ in mJy arcsec$^{-2}$ and $d$ in kpc.  $I_0$
can be measured from the contour levels of a radio image (for
significantly extended sources) or can be obtained by dividing the
source total flux by the source solid angle, while $d$ can be inferred from
geometrical arguments.  The constant has been computed  for $\alpha$ =
0.7, $\nu_1$ = 10 MHz and $\nu_2$ = 100 GHz (tabulated 
in \cite{Gov04a}, for other values of these parameters).

The magnetic field for which the total energy content is minimum is
referred to as the equipartition value and is derived as follows:
\begin{equation}
B_{eq} = \left({{24\pi}\over{7}} u_{\rm min}\right)^{\frac{1}{2}}.
\label{beqfin}
\end{equation}
\noindent
One must be aware of the uncertainties inherent to the determination
of the minimum energy density and equipartition magnetic field
strength.  The value of $k$, the ratio of the energy in relativistic
protons to that in electrons (eqn. \ref{kpar}), depends on the
mechanism of generation of relativistic electrons, which, so far,
is poorly known.  Values usually assumed in literature for clusters are
$k$ = 1 (or $k$ = 0).  Uncertainties are also related to the volume
filling factor $\Phi$.

\begin{figure}[t]
\centering
\includegraphics[height=8cm]{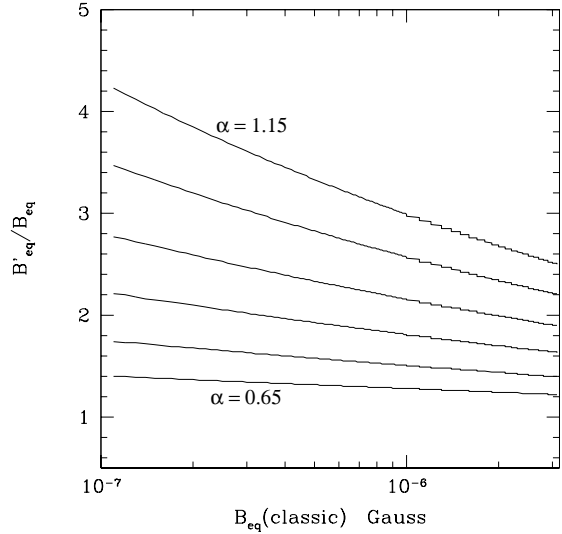}
\caption{
Values of the ratio B$^{\prime}_{eq}$/B$_{eq}$ (see text) as a function of 
the equipartition magnetic field obtained with the classical approach,
assuming an electron minimum Lorentz factor $\gamma_{\rm min}$ = 50.
Different lines refer to different values of the initial spectral index
(i.e. not affected by ageing),
from $\alpha$ = 1.15 (top line) to  $\alpha$ = 0.65 (bottom line) in
steps of  $\alpha$ = 0.1.
}
\label{fig3}       % Give a unique label
\end{figure}

In the standard approach presented above, the equipartition parameters
are obtained from the synchrotron radio luminosity observed between
the two fixed frequencies $\nu_1$ and $\nu_2$. Brunetti et al. \cite{Bru97}
demonstrated that it is more appropriate to calculate the radio source
energy by integrating the synchrotron luminosity over a range of
electron energies.  This avoids the problem that electron energies
corresponding to frequencies $\nu_1$ and $\nu_2$ depend on the
magnetic field value (see eqn. \ref{sync}), thus the integration over a
range of fixed frequencies is equivalent to considering radiating electrons
over a variable range of energies. Moreover, it has the advantage that
electrons of very low energy are also taken into account. The
equipartition quantities obtained by following this approach are
presented by \cite{Bru97} and \cite{Bec05}.
Representing the electron energy by its Lorentz factor $\gamma$,
and assuming that $\gamma_{\rm min} \ll \gamma_{\rm max}$, the new 
expression for the equipartition magnetic field B$^{\prime}_{eq}$ in Gauss (for
$\alpha~>~0.5$) is:

\begin{equation}
B^{\prime}_{eq} \sim 1.1 ~ \gamma_{\rm min}^{{1-2\alpha}\over{3+\alpha}}~B_{eq}^{{7}\over{2(3+\alpha)}},
\label{eqbru}
\end{equation}

\noindent
where B$_{eq}$ is the value of the equipartition magnetic field
obtained with the standard formulae by integrating the radio spectrum
between 10 MHz and 100 GHz.
It should be noticed that  B$^{\prime}_{eq}$ is larger than B$_{eq}$
for B$_{eq}$ $<$ $\gamma_{\rm min}^{-2}$ (see Fig. \ref{fig3}). 

\section{Radio emission from the ICM: diffuse radio sources}
\label{s:diff}

\begin{figure}[t]
\centering
\includegraphics[width=6.5cm]{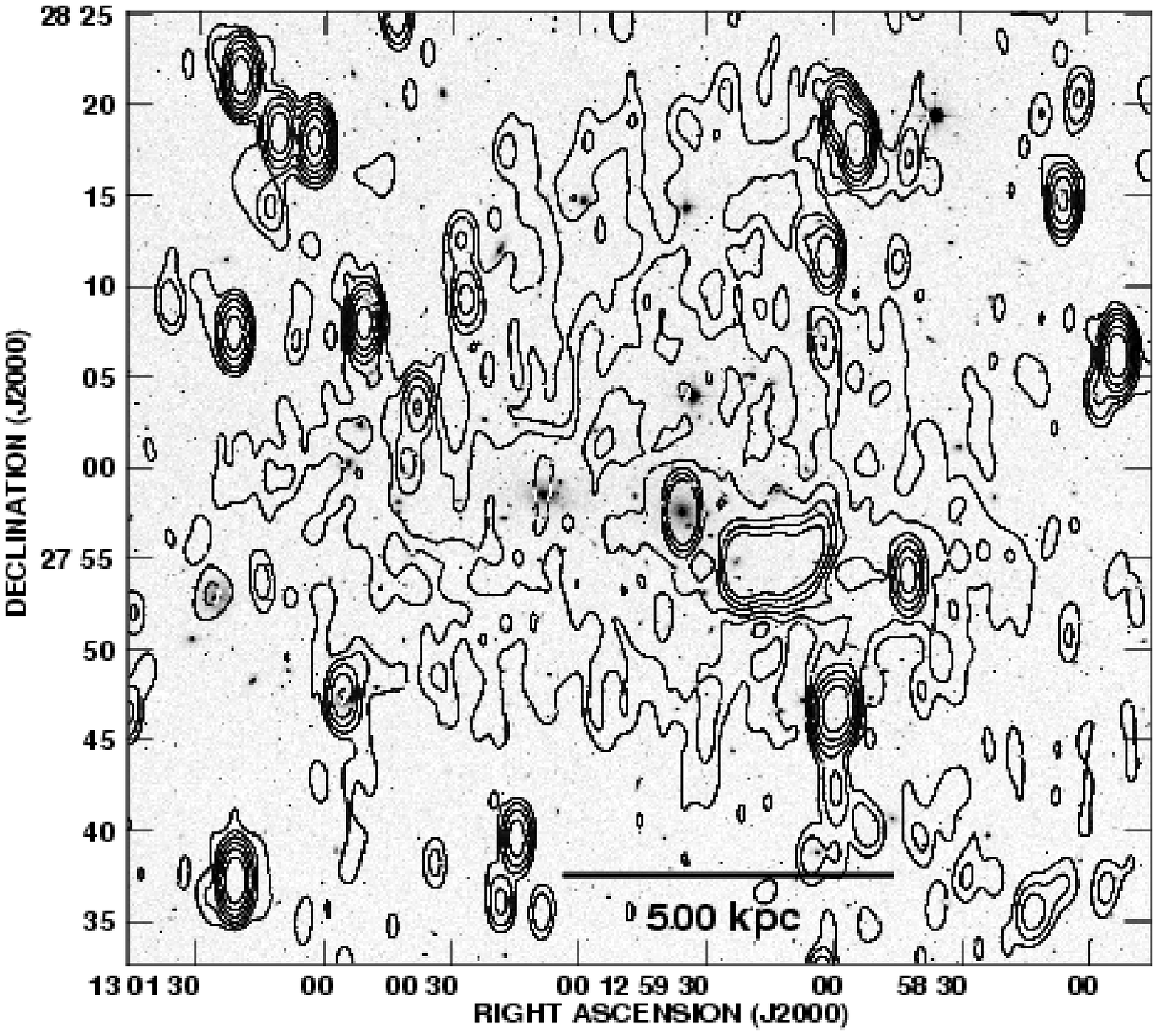}
%\includegraphics[width=6.5cm]{Fer_fig4.ps}
%\hspace{2in}
\includegraphics[width=5.0cm]{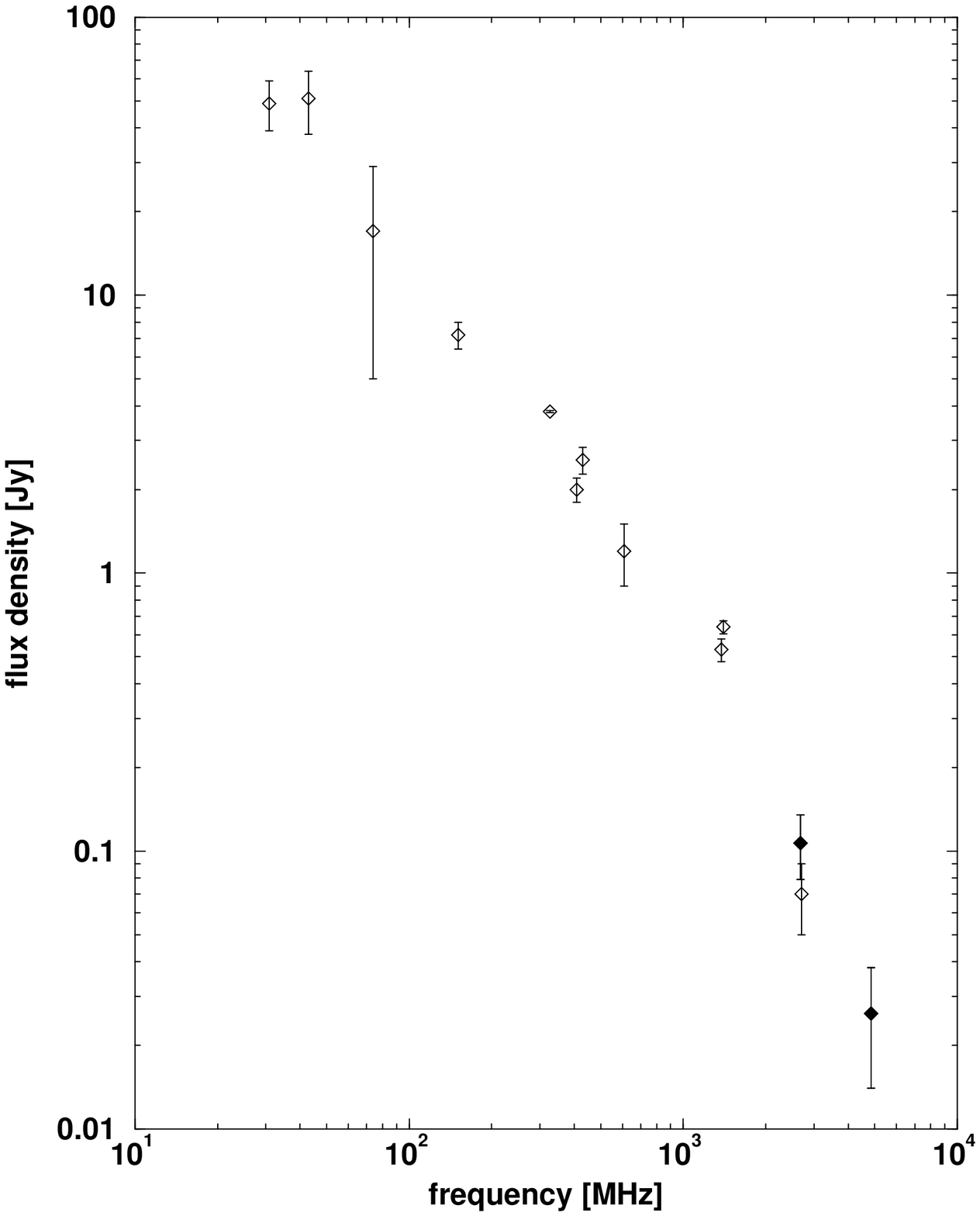}
\caption{
{\bf Left panel:} Diffuse radio halo Coma C in the Coma cluster ($z=0.023$) at 
 0.3 GHz, superimposed onto the optical image from the DSS1.
The resolution of the radio image is 55\arcsec$\times$ 125\arcsec
(FWHM, RA $\times$ DEC); contour levels are: 3, 6, 12, 25, 50, 100 mJy/beam.
{\bf Right panel:} Total radio spectrum of the radio halo Coma C
(from \cite{Thi03}).
}
\label{coma}       % Give a unique label
\end{figure}

In recent years, there has been growing evidence for the existence of
cluster large-scale diffuse radio sources, which have no optical
counterpart and no obvious connection to cluster galaxies, and are
therefore associated with the ICM.  These sources are typically
grouped in 3 classes: halos, relics and mini-halos.  The number of
clusters with halos and relics is presently around 50, and whose 
properties have been recently reviewed by Giovannini \& Feretti \cite{Gio02}
and Feretti \cite{Fer03}.  The synchrotron nature of this radio emission
indicates the presence of cluster-wide magnetic fields of the order of
$\sim$ 0.1-1 $\mu$G, and of a population of relativistic electrons
with Lorentz factor $\gamma \gg$ 1000. The understanding of these
non-thermal components is important for a comprehensive physical
description of the ICM.

\subsection{Radio halos}
\label{s:halos}

\begin{figure}[t]
\centering
\includegraphics[height=5.0cm]{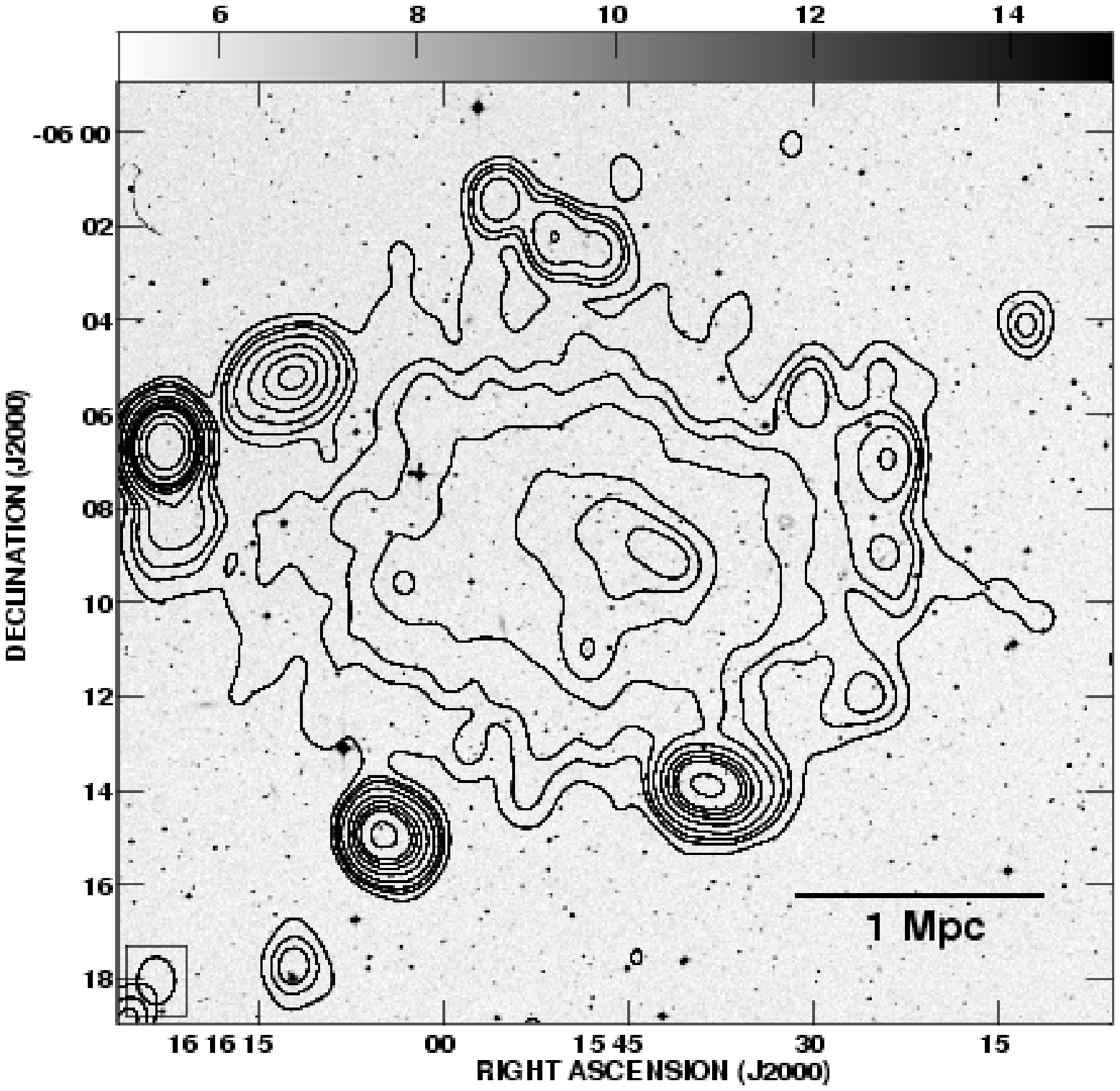}
%\includegraphics[height=5.5cm]{Fer_fig5.ps}
%\hspace{2in}*
\includegraphics[height=5.0cm]{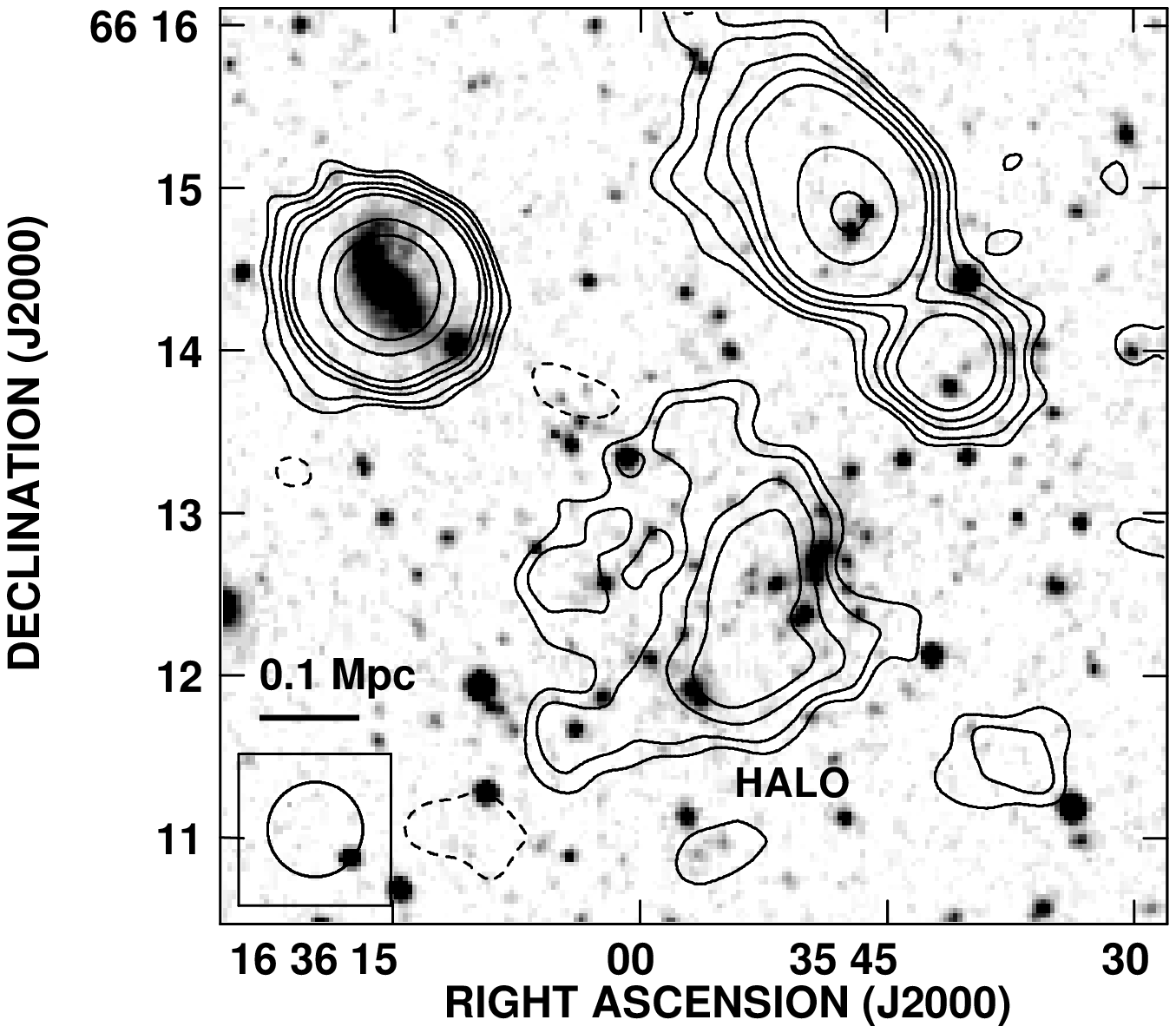}
\caption{
{\bf Left panel:} Radio emission in A2163 ($z=0.203$) at 20 cm
\cite{Fer01}. 
The radio halo is one of the most powerful and extended halos
known so far.
{\bf Right panel:} 
Radio emission  of the cluster A2218 ($z=0.171$) at 20 cm 
\cite{Gio00}. In both clusters the radio contours
are overlayed onto the grey-scale optical image.
}
\label{aloni}       % Give a unique label
\end{figure}

Radio halos are diffuse radio sources of low surface brightness
($\sim$ $\mu$Jy arcsec$^{-2}$ at~20~cm) permeating the central volume
of a cluster. They are typically extended with sizes \gtsim~1 Mpc and
are unpolarized down to a few percent level.  The prototype of this
class is the diffuse source Coma C at the centre of the Coma cluster
(\cite{Gio93} and Fig. \ref{coma}), first classified by
Willson \cite{Wil70}.  The halo in A2163, shown in left panel of
Fig. \ref{aloni}, is one of the most extended and powerful halos.
Other well studied giant radio halos are present in A665 \cite{Gio00}, 
A2219 \cite{Bac03}, A2255 \cite{Fer97a}, A2319 \cite{Fer97b}, A2744 (Fig. \ref{relitti},
left panel), 1E0657-56 \cite{Lia00}, and in the distant cluster
CL 0016+16 \cite{Gio00} at redshift $z=0.555$.  All
these clusters show recent merging processes, and no cooling core.

Radio halos of small size, i.e. $\ll$ 1 Mpc, have also been revealed in
the central regions of clusters.  Some examples are in A401
\cite{Gio00}, A1300 \cite{Reid99}, A2218
(Fig. \ref{aloni}, right panel) and A3562 \cite{Gia05}.
All these clusters, as well as those hosting giant radio halos, are
characterized by recent merger processes and no cooling core.

Unlike the presence of thermal X-ray emission, the presence of diffuse
radio emission is not common in clusters of galaxies: the detection
rate of radio halos, at the detection limit of the NRAO VLA Sky Survey
(NVSS) is $\sim$ 5\% in a complete cluster sample \cite{Gio99}.  
However, the probability is much larger, if clusters with
high X-ray luminosity are considered.  Indeed, $\sim$ 35\% of clusters
with X-ray luminosity larger than 10$^{45}$ erg s$^{-1}$ X-ray (in the
ROSAT band 0.1-2.4 keV, computed assuming $H_0$ = 50 km
s$^{-1}$Mpc$^{-1}$ and q$_0$ = 0.5) show a giant radio halo
\cite{Gio02}.

The physical parameters in radio halos can be estimated assuming
equipartition conditions, and further assuming equal energy in
relativistic protons and electrons, a volume filling factor of 1, a
low frequency cut-off of 10 MHz, and a high frequency cut-off of 10
GHz.  The derived minimum energy densities in halos and relics are of
the order of 10$^{-14}$ - 10$^{-13}$ erg cm$^{-3}$, i.e.  much lower
than the energy density in the thermal gas.  The corresponding
equipartition magnetic field strengths range from 0.1 to 1 $\mu$G.

The total radio spectra of halos are steep
($\alpha$\gtsim~1)\footnote{S($\nu$) 
$\propto$ $\nu^{-\alpha}$ as in eqn. \ref{spectralind}}, as
typically found in aged radio sources.  Only a few halos have good
multi-frequency observations that allow an accurate determination of
their integrated spectrum.  Among them, the spectrum of the Coma cluster
halo is characterized by a steepening at high frequencies, which has
been recently confirmed by single dish data (Fig. \ref{coma}, right
panel). The spectrum of the radio halo in A1914 is very steep, with an
overall slope of $\alpha \sim$ 1.8. A possible high frequency
curvature is discussed by Komissarov \& Gubanov \cite{Kom94}.  In A754,
Bacchi et al. \cite{Bac03} e
stimate $\alpha_{\rm 0.07GHz }^{\rm 0.3GHz}$ $\sim$ 1.1, and
$\alpha_{\rm 0.3GHz}^{\rm 1.4GHz}$ $\sim$ 1.5, 
and infer the presence of a possible
spectral cutoff.  Indication of a high frequency spectral steepening
is also obtained in the halo of A2319, where Feretti et al. \cite{Fer97b}
report $\alpha_{\rm 0.4GHz}^{\rm 0.6GHz}$ $\sim$ 0.9 and 
$\alpha_{\rm 0.6GHz}^{\rm 1.4GHz}$ $\sim$ 2.2.
In the few clusters where maps of the spectral index are available (Coma
C, \cite{Gio93}; A665 and A2163, \cite{Fer04}, the
radio spectrum steepens radially with the distance from the cluster
centre.  In addition, it is found that the spectrum in A665 and A2163
is flatter in the regions influenced by merger processes (see 
\S~\ref{s:connect}).

\begin{figure}[t]
\centering
\includegraphics[width=5.5cm]{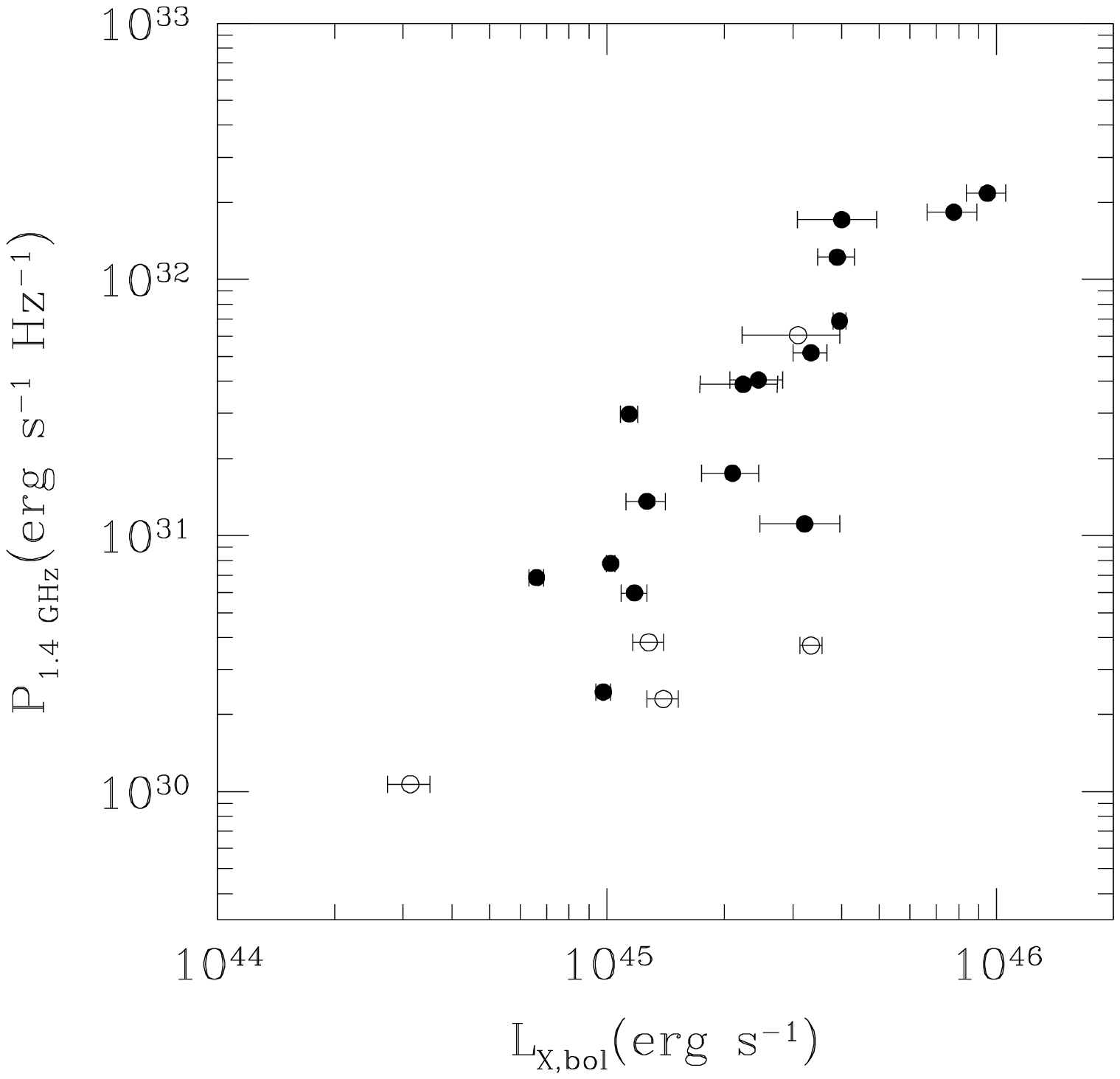}
\includegraphics[width=5.5cm]{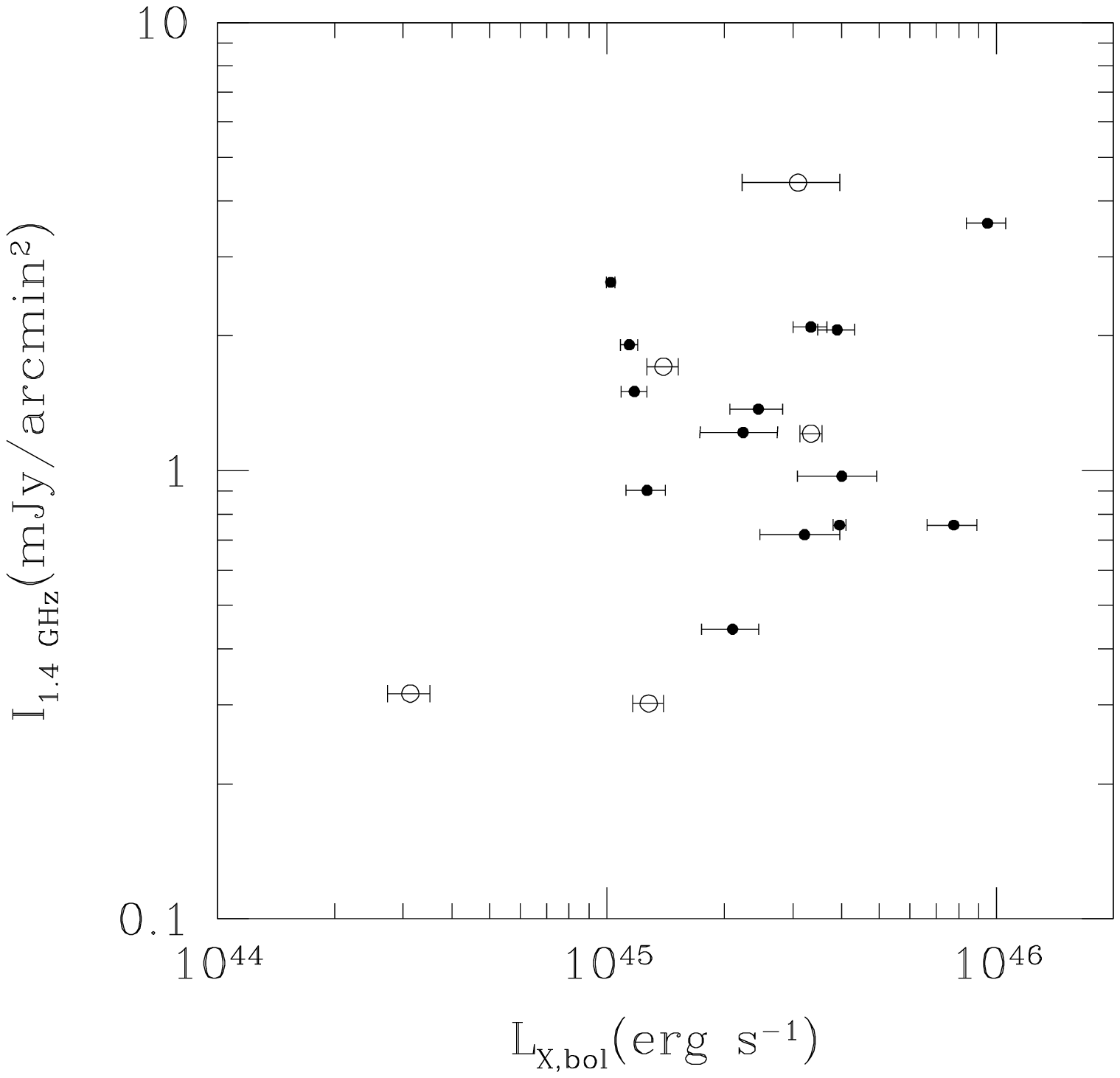}
\caption{ {\bf Left panel:} Monochromatic radio power at 20 cm versus
cluster bolometric X-ray luminosity.  {\bf Right panel:} Average
surface brightness of the radio halos versus cluster
X-ray luminosity.  In both
panels, filled and open circles refer to halos of size $>$ and $<$ 1 Mpc,
respectively. }
\label{correl}
\end{figure}

In general, from the spectra of halos, it is derived
that the radiative lifetime of the relativistic electrons, considering
synchrotron and inverse Compton energy losses, is of the order of
$\sim$ 10$^8$ yr \cite{Sar99}.  
Since the 
expected diffusion velocity of the electron population is of the order
of the Alfv\'en speed ($\sim$ 100 km s$^{-1}$), the radiative electron 
lifetime is too short to allow the particle diffusion
throughout the cluster volume. Thus, the radiating electrons cannot
have been produced at some localized point of the cluster, but they
must undergo {\it in situ} energization, acting with an efficiency comparable
to the energy loss processes \cite{Pet01}.  We will show in 
\S~\ref{s:partic} that recent cluster mergers are likely to supply 
energy to the halos and relics.  

The radio and X-ray properties of halo clusters are related.  The most
powerful radio halos are detected in the clusters with the highest
X-ray luminosity. This follows from the correlation shown in
Fig. \ref{correl} between the monochromatic radio power of a halo at
20 cm and the bolometric X-ray luminosity of the parent cluster
\cite{Lia00, Gio02}. The right panel of
Fig. \ref{correl} shows the correlation between the average surface
brightness of the radio halo and the cluster X-ray luminosity. Since
the brightness is an observable, this correlation can be used to
set upper limits to the radio emission to those clusters in which a
radio halo is not detected.  It is worth reminding the reader that the radio
power versus X-ray luminosity correlation is valid for merging clusters
with radio halos, and therefore cannot be generalized to all clusters.
Among the clusters with high X-ray luminosity and no radio halo, there
are A478, A576, A2204, A1795, A2029, all well known relaxed clusters
with a massive cooling flow.  An extrapolation of the above
correlation to low radio and X-ray luminosities indicates that
clusters with $L_{X}$ \ltsim~10$^{45}$ erg s$^{-1}$ would host halos
of power of a few 10$^{23}$ W Hz$^{-1}$.  With a typical size of 1
Mpc, they would have a radio surface brightness 
(easily derived from the right panel of Fig. \ref{correl})
lower than current limits obtained in the literature and in the NVSS. 
On the other hand,
it is possible that giant halos are only present in the most X-ray
luminous clusters, i.e. above a threshold of X-ray luminosity (see
\cite{Bac03}).  Future radio data with next generation
instruments (LOFAR, LWA, SKA) will allow the detection of low
brightness/low power large halos, in order to clarify if halos are
present in all merging clusters or only in the most massive ones.

Since cluster X-ray luminosity and mass are correlated \cite{Reip02}, 
the correlation between radio power ($P_{\rm
1.4~GHz}$) and X-ray luminosity could reflect a dependence of the
radio power on the cluster mass. A correlation of the type $P_{\rm
1.4~GHz}$ $\propto$ $M^{2.3}$ has been derived \cite{Gov01a},
\cite{Fer03}, where $M$ is the total gravitational mass within a
radius of 3$ h^{-1}_{50}$ Mpc.
%($H_0$/50)$^{-1}$ Mpc.
Using the cluster mass within the virial radius, the correlation is
steeper (Cassano et al. in preparation).  A correlation of radio power
vs cluster mass could indicate that the cluster mass may be a crucial
parameter in the formation of radio halos, as also suggested by
\cite{Buo01}.  
Since it is likely that massive clusters are the result of
several major mergers, it is  concluded that both past mergers and current
mergers are the necessary ingredients for the formation and evolution
of radio halos. This scenario may provide a further explanation of the
fact that not all clusters showing recent mergers host radio
halos, which is expected from the recent modeling of Cassano \&
Brunetti \cite{Cas05}.

\begin{figure}[t]
\centering
\includegraphics[width=5.5cm]{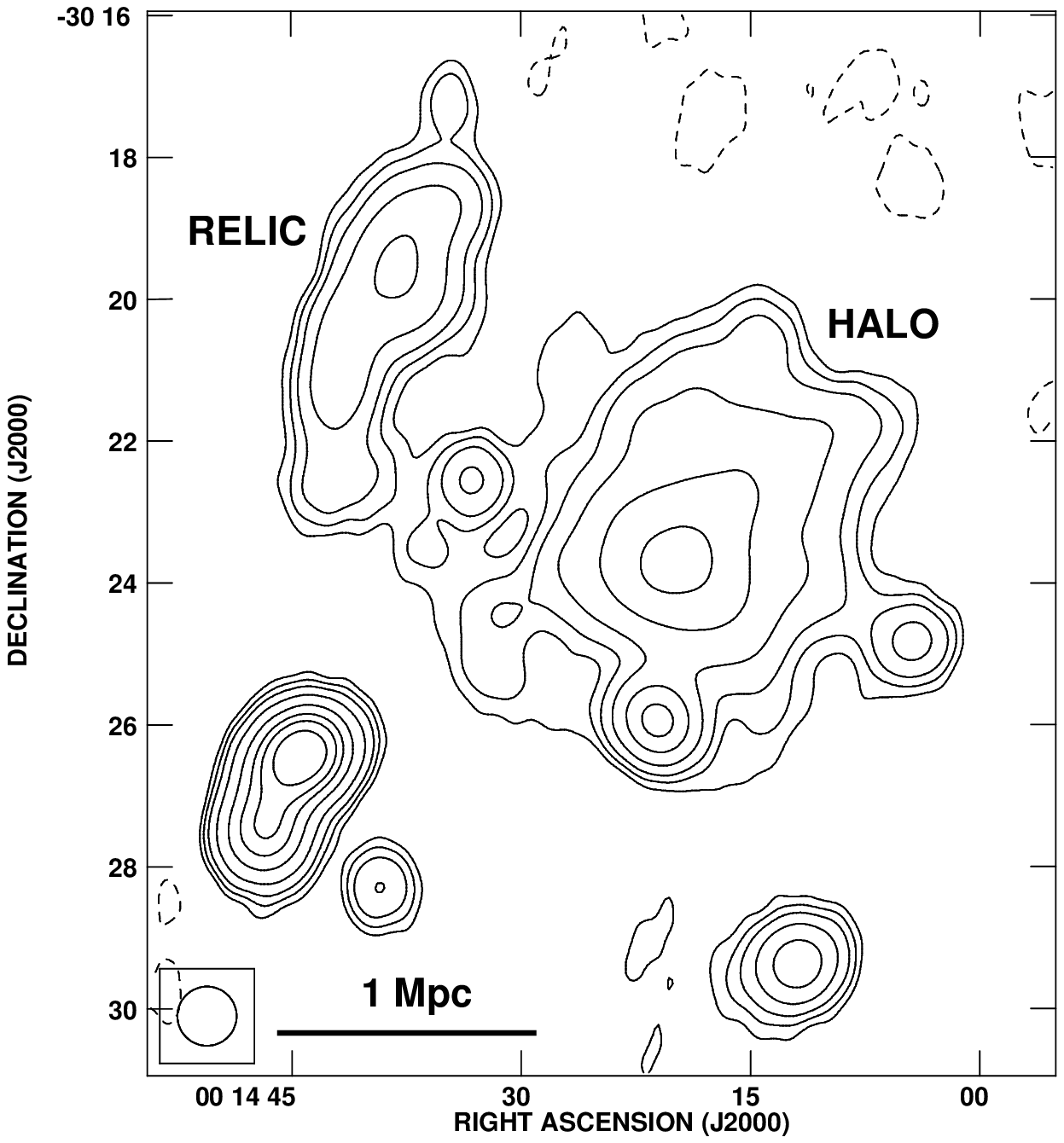}
\includegraphics[width=6.0cm]{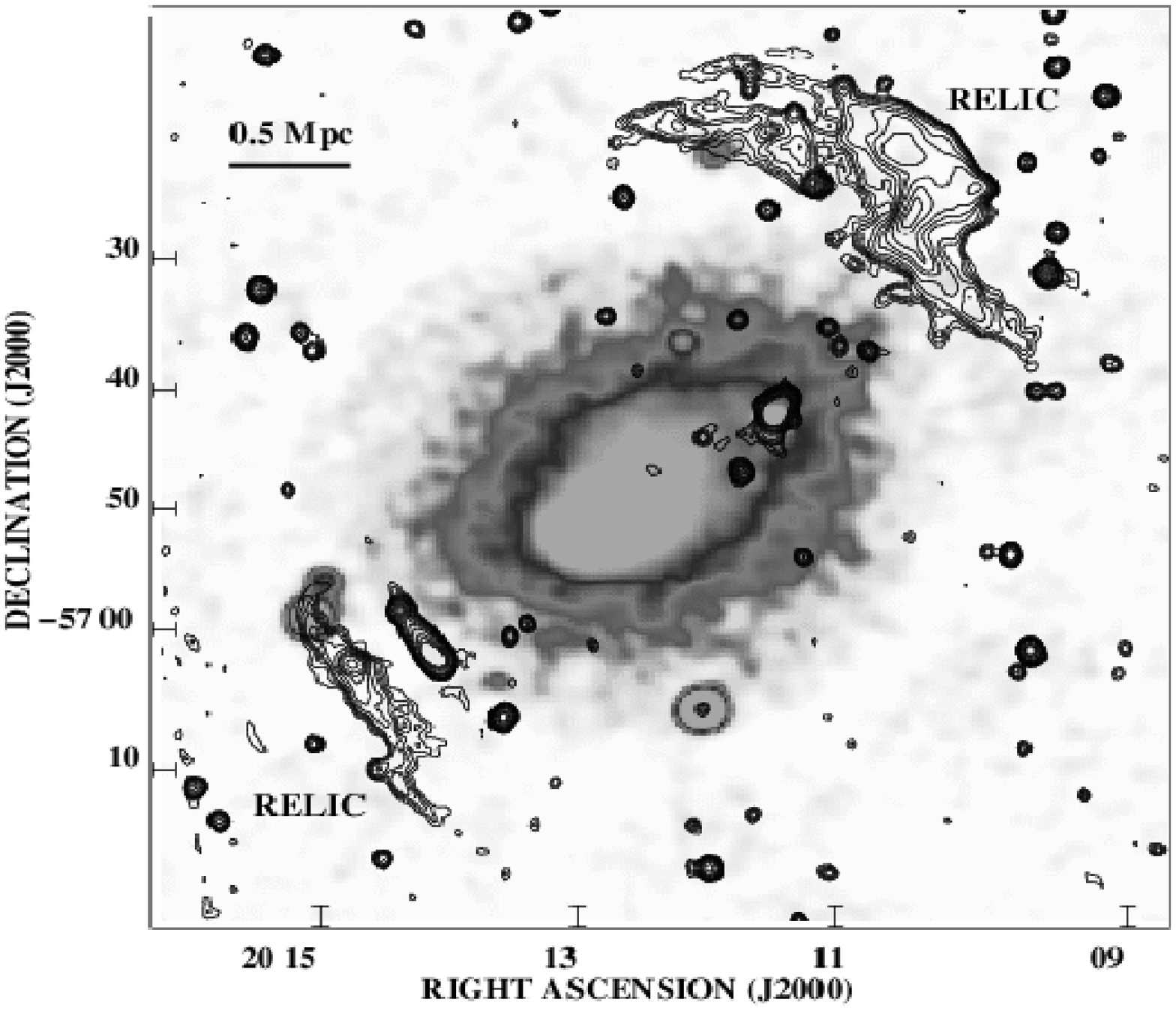}
\caption{
{\bf Left panel:} Radio emission of  A2744  ($z=0.308$) showing a peripheral
elongated relic, and a central radio halo \cite{Gov01a}. 
{\bf Right panel:} A3667 ($z=0.055$): contours
of the radio emission at 36 cm \cite{Rot97} overlayed onto the grey-scale ROSAT X-ray image. 
Two radio relics are located on opposite sides of the
cluster along the axis of the merger, with the individual radio
structures elongated perpendicular to this axis.
}
\label{relitti}
\end{figure}

\subsection{Radio relics}
\label{s:relics}

Relic sources are diffuse extended sources, similar to the radio halos
in their low surface brightness, large size (\gtsim~1 Mpc) and steep
spectrum ($\alpha$ \gtsim~ 1), but they are generally detected in the
cluster peripheral regions.  They typically show an elongated radio
structure with the major axis roughly perpendicular to the direction
of the cluster radius, and they are strongly polarized ($\sim$
20-30\%).  The most extended and powerful sources of this class are
detected in clusters with central radio halos: in the Coma cluster
(the prototype relic source 1253+275, \cite{Gio91}, A2163
\cite{Fer01}, A2255 \cite{Fer97a}, A2256
\cite{Rot94} and A2744 (Fig. \ref{relitti}, left panel).
A spectacular example of two giant almost symmetric
relics in the same cluster is found in A3667 (Fig. \ref{relitti},
right panel). There are presently only a few cases of double opposite 
relics in clusters.

\begin{figure}[t]
\centering
\includegraphics[width=5.5cm]{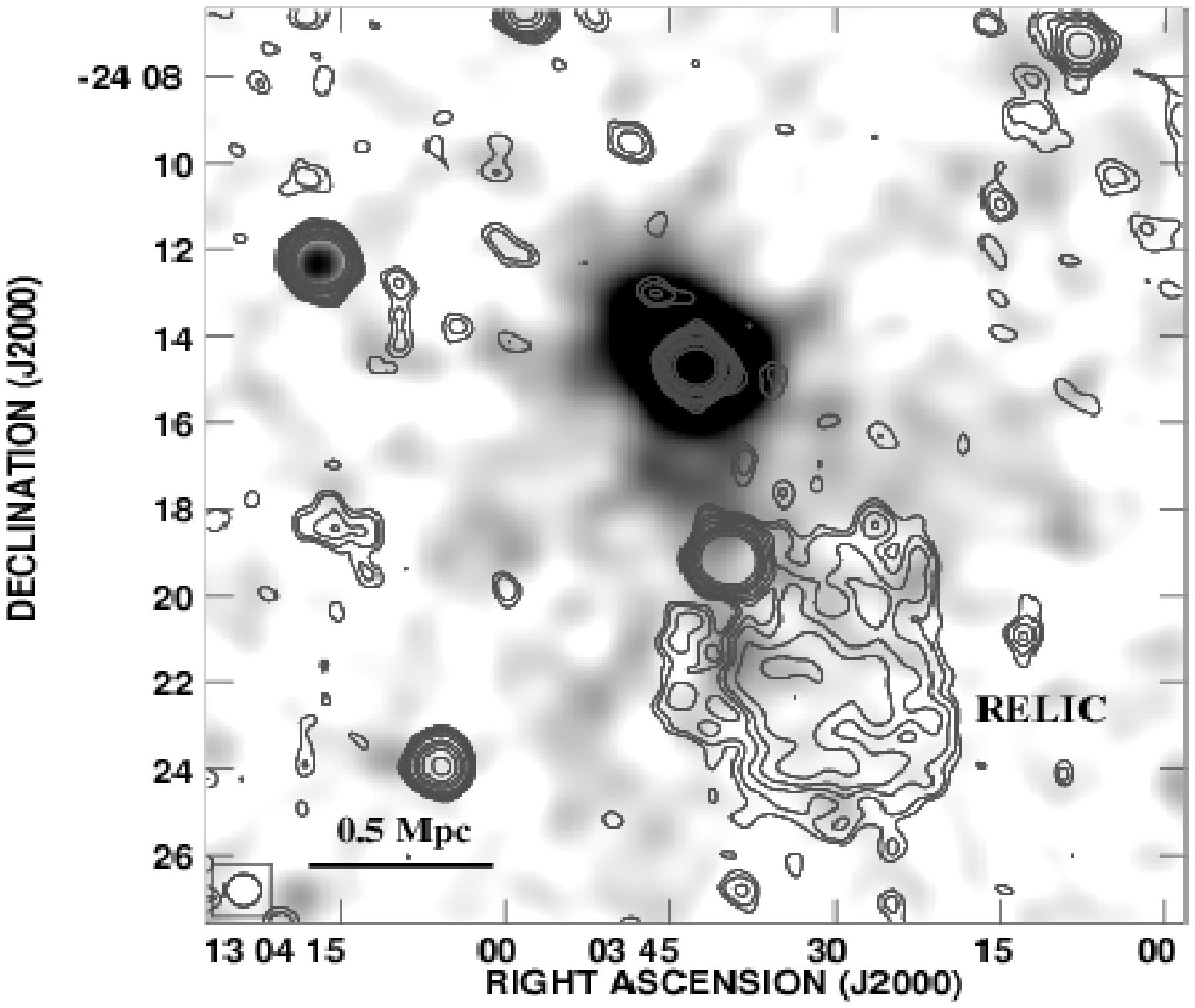}
\includegraphics[width=6.0cm]{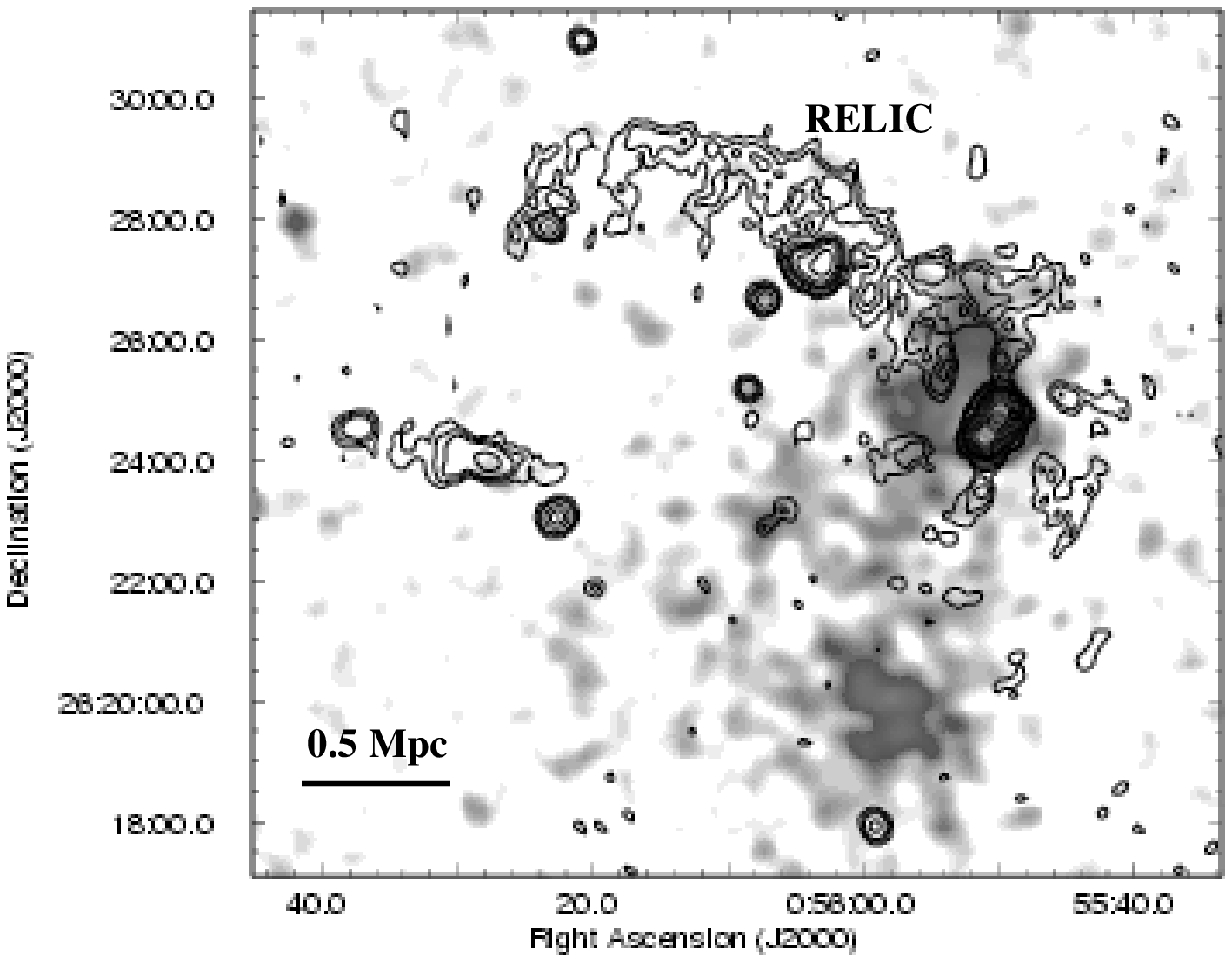}
\caption{
Radio emission at 20 cm (contours) of the clusters: 
{\bf Left panel:} A1664 ($z=0.128$), {\bf Right panel:} A115 ($z=0.197$), 
superimposed onto the grey-scale cluster 
X-ray emission detected from ROSAT PSPC \cite{Gov01a}.
}
\label{relitti2}
\end{figure}

Other morphologies have been found to be associated with relics (see
\cite{Gio04} for a review). In the cluster A1664
(Fig. \ref{relitti2}, left panel), the structure is approximately
circular and
regular. In A115 (Fig. \ref{relitti2}, right panel), the elongated
relic extends from the cluster center to the periphery. This could be
due to projection effects, however this is the only relic showing such
behaviour.

\begin{figure}
\centering
\includegraphics[height=6.5cm]{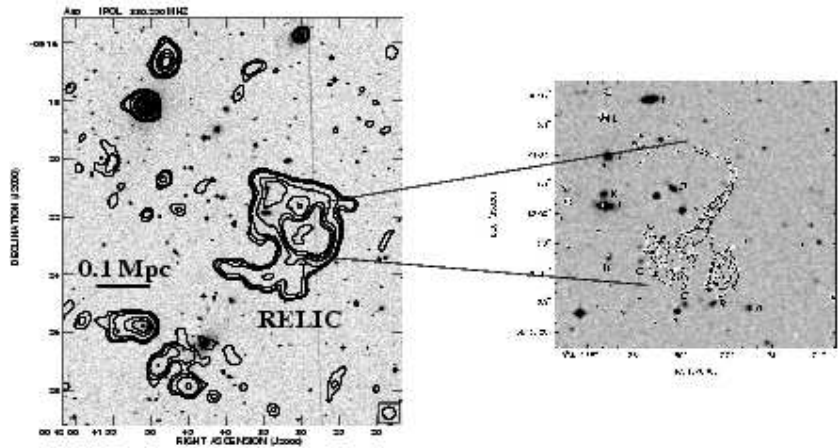}
\caption{
Radio emission at 90 cm (contours) in A85 ($z=0.056$), 
superimposed onto the opical image \cite{Gio00}.
The zoom to the right shows the filamentary structure detected at 
high resolution by Slee et al. \cite{Sle01} at 20 cm.}
\label{a85}
\end{figure}

There are diffuse radio sources which are naturally classified as
relics, because of their non-central cluster location, but their
characteristics are quite different from those of giant relics.  
Examples of these sources are in A13, A85 (Fig. \ref{a85}), A133,
A4038 \cite{Sle01}: they show a much smaller size than relics
(\ltsim~300 kpc down to $\sim$ 50 kpc), are generally closer to the
cluster center, and show extremely steep radio spectra ($\alpha$
\gtsim~2). They are strongly polarized (\gtsim~30\%), and often
quite filamentary  when observed with sufficient resolution.
The relic in A133 was suggested to be related to past 
activity of a nearby galaxy \cite{Fuj02}.

The detection rate of radio relics in a complete sample 
of clusters is $\sim$ 6\% at the detection limit of the
NVSS \cite{Gio02}.
Relics are found in clusters both with and without a cooling
core, suggesting that they may be related to minor or off-axis
mergers, as well as to major mergers.  Theoretical models propose that
they are tracers of shock waves in merger events (see
\S~\ref{s:origre}). This is consistent with their elongated
structure, almost perpendicular to the merger axis.
%The production of
%outgoing shock waves at the cluster periphery is indeed observed in
%numerical simulations of cluster merger events (Ryu et al. 2003).  
The radio power of relics correlates with the cluster X-ray luminosity
(\cite{Fer02a, Gio04}, as also found for halos (see
\S~\ref{s:halos} and Fig. \ref{correl}), although with a larger
dispersion.  The existence of this correlation indicates a link
between the thermal and relativistic plasma also in peripheral cluster
regions.

\begin{figure}
\centering
\includegraphics[width=5.5cm]{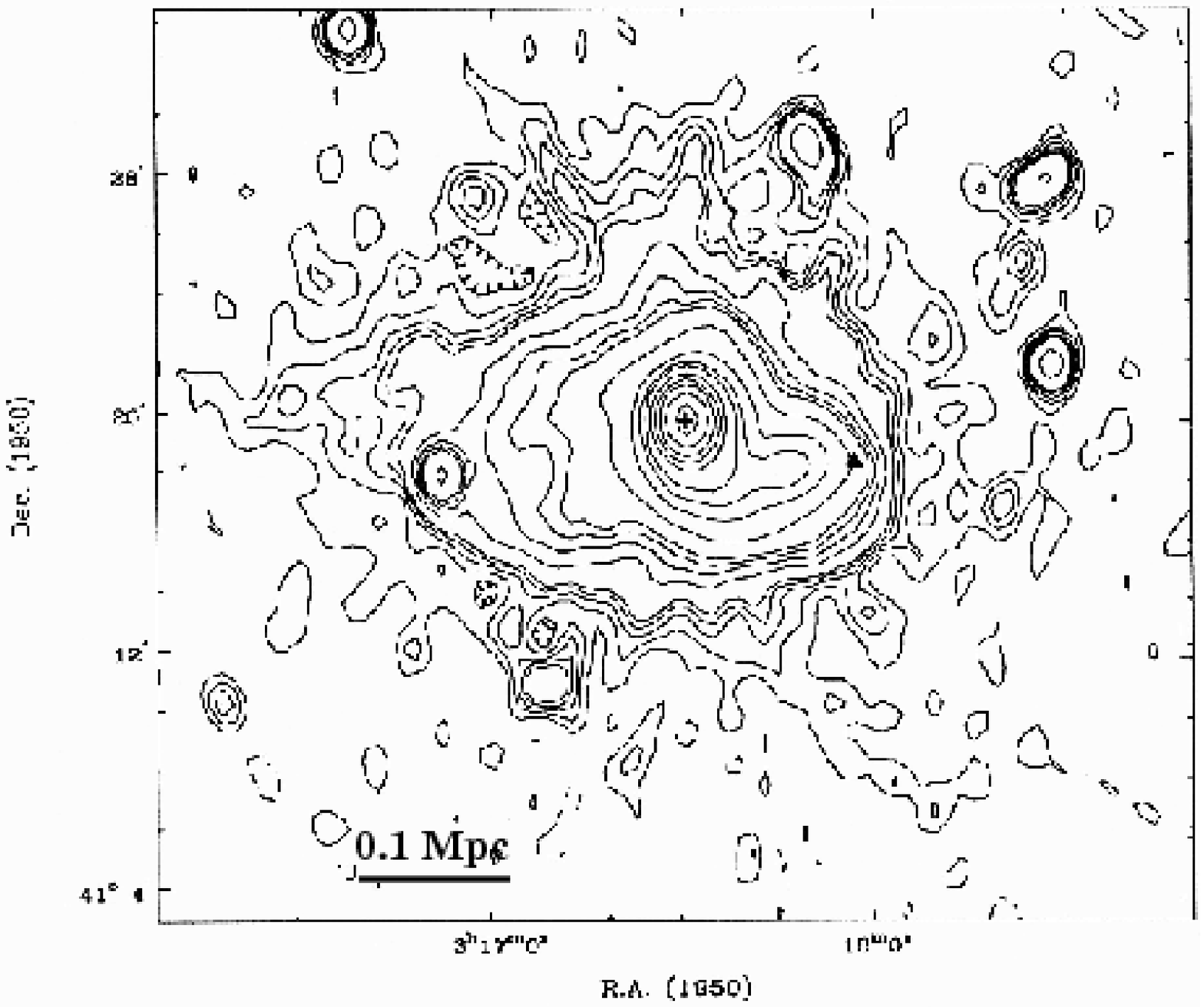}
\includegraphics[width=5.5cm]{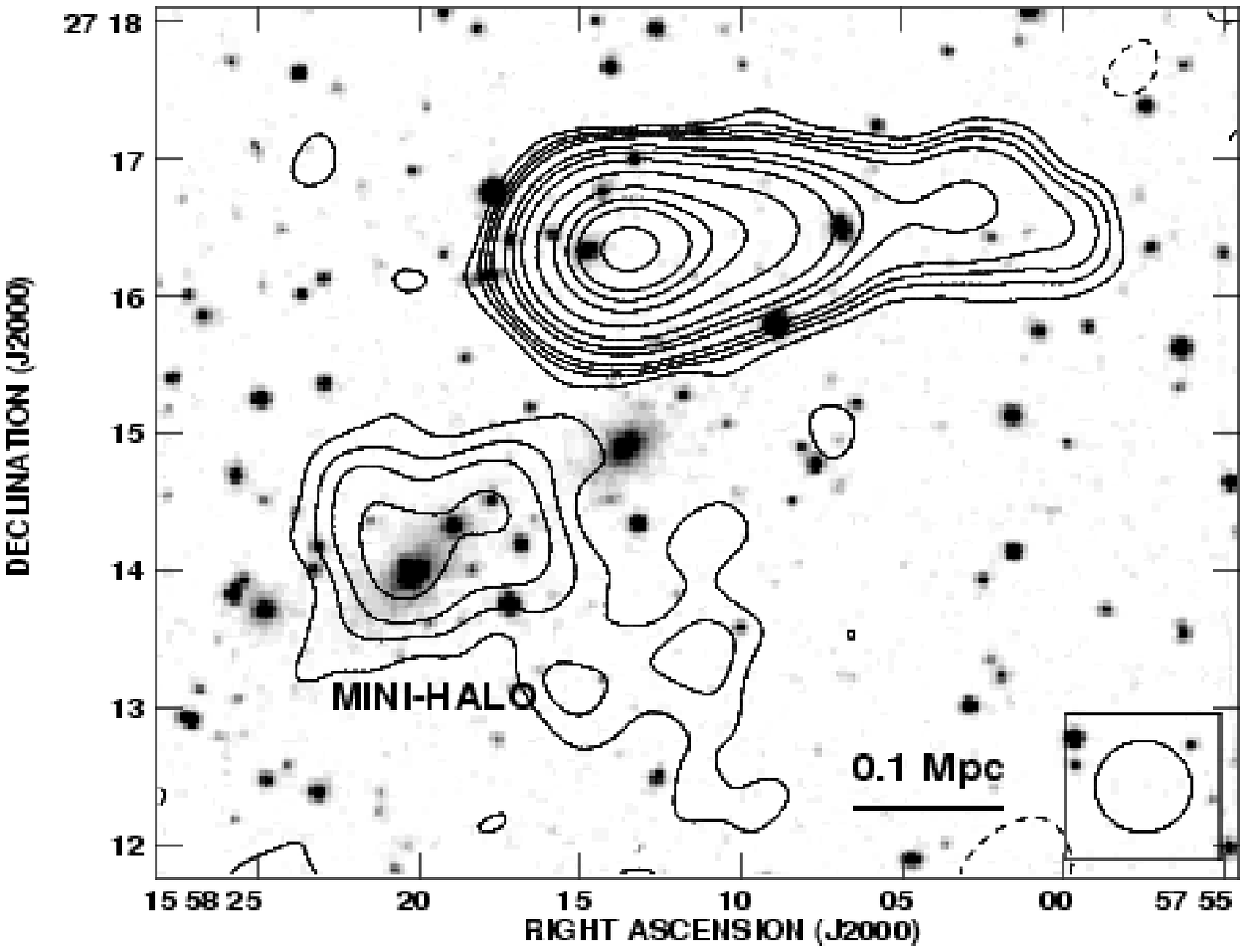}
\caption{
{\bf Left panel:} Radio contour map of the mini-halo in the Perseus 
cluster ($z=0.018$) at 92 cm \cite{Sij93},
{\bf Right panel:} The mini-halo in A2142 ($z=0.089$), 
superimposed onto the optical image \cite{Gio00}.
}
\label{minih}
\end{figure}

\subsection{Mini-halos}
\label{s:minih}

Mini-halos are small size ($\sim$ 500 kpc) diffuse radio sources at
the center of cooling core clusters, usually surrounding a powerful
radio galaxy, as in the Perseus cluster (Fig. \ref{minih}, left
panel), Virgo cluster \cite{Owe00}, PKS 0745-191 \cite{Bau91}, 
A2626 \cite{Git04}.  Since there is an anticorrelation
between the presence of a cooling core and that of a major merger
event, mini-halos are the only diffuse sources which are not
associated with cluster mergers.  A peculiar example is represented by
the cluster A2142, which contains a cooling core but also shows a cold
front and thus merging activity \cite{Mar00}. The
mini-halo in this cluster is about 200 kpc in size and does not
surround any powerful radio galaxy (Fig. \ref{minih}, right panel).
For the latter reason, it could be also considered as a small halo.

The radio spectra of mini-halos are steep, as those of halos and
relics.  In the Perseus mini-halo, the integrated spectrum steepens at
high frequency and the spectral index distribution shows a radial
steepening \cite{Sij93}.

Gitti et al. \cite{Git02} argued that the radio emitting particles in
mini-halos cannot be connected to the central radio galaxy in terms of
particle diffusion or buoyancy, but they are likely associated with
the ICM in the cooling flow region (see \S~\ref{s:origmh}).  This
is supported by the correlation observed between the mini-halo radio
power and the cooling flow power \cite{Git04}; however, the
number of objects is still low and the parameters are affected by
large errors.

\section{Radio emitting particles}
\label{s:partic}

>From the diffuse radio emission described in the previous sections, it
is determined that highly energetic relativistic electrons ($\gamma \sim$
10$^4$) are present in clusters, either in the central or in the
peripheral regions.  They are found both in merging (halos and
relics) and relaxed (mini-halos) clusters, thus under different
cluster conditions.  These radio features are currently not known to
be present in all clusters.  They show steep radio spectra, thus the
radiating particles have short lifetimes ($\sim$ 10$^8$ yr).  Given the
large size of the radio emitting regions, the relativistic particles
need to be reaccelerated by some mechanism, acting with an efficiency
comparable to the energy loss processes.  Several possibilities have
been suggested for the origin of relativistic electrons and for the
mechanisms transferring energy into the relativistic electron
population.

\subsection{Connection between halos/relics and cluster merger processes}
\label{s:connect}

Evidence favour the hypothesis that clusters with halos and
relics are characterized by strong dynamical activity, related to
merging processes.  These clusters indeed show: (i) substructures and
distortions in the X-ray brightness distribution \cite{Sch01}; 
(ii) temperature gradients \cite{Mar98} and gas
shocks \cite{Mar03a}; (iii) absence of a strong cooling
flow \cite{Sch01}; (iv) values of the spectroscopic $\beta$
parameter which are on
average larger than 1 \cite{Fer02a}; (v) core radii significantly
larger than those of clusters classified as single/primary
\cite{Fer02a}; 
(vi) larger distance from the nearest neighbours, compared to
clusters with similar X-ray luminosity \cite{Sch99}.  
The fact that they appear more isolated supports the idea that
recent merger events lead to a depletion of the nearest neighbours.

Buote \cite{Buo01} derived a correlation between the radio power of halos and
relics and the dipole power ratio of the cluster two-dimensional
gravitational potential.  Since power ratios are closely related
to the dynamical state of a cluster, this correlation represents
the first attempt to quantify the link between diffuse sources and
cluster mergers.

Maps of the radio spectral index between 0.3 GHz and 1.4 GHz of the
halos in the two clusters A665 and A2163 show that the regions influenced
by the merger, as deduced from X-ray maps, show flatter spectra
\cite{Fer04}.  This is the first direct confirmation that the
cluster merger supplies energy to the radio halo.  Finally, we point
out that we are not presently aware of any radio halo or relic in a
cluster where the presence of a merger has been clearly excluded.

\subsection{Relativistic electrons in radio halos}
\label{s:origha}

\noindent
{\bf Origin}. The relativistic electrons present in the cluster
volume, which are responsible for the diffuse radio emission, can be either {\it
primary } or {\it secondary electrons}. Primary electrons were
injected into the cluster volume by AGN activity (quasars, radio
galaxies, etc.), or by star formation in normal galaxies (supernovae,
galactic winds, etc.) during the cluster dynamical history.  This
population of electrons suffers strong radiation losses mainly because
of synchrotron and inverse Compton emission, thus reacceleration is
needed to maintain their energy to the level necessary to produce
radio emission.  For this reason, primary electron models can also be
referred to as reacceleration models.
These models predict that the accelerated electrons have a maximum
energy at $\gamma <$ 10$^5$ which produces a high frequency cut-off in
the resulting synchrotron spectrum \cite{Bru04a}.  Thus a high
frequency steepening of the integrated spectrum is expected, as well
as a radial steepening and/or a complex spatial distribution of the
spectral index between two frequencies, the latter due to different
reacceleration processes in different cluster regions.  Moreover, in
these models, a tight connection between radio halos and cluster
mergers is expected.

Secondary electrons are produced from inelastic nuclear collisions
between the relativistic protons and the thermal ions of the ambient
intracluster medium.  The protons diffuse on large scales because
their energy losses are negligible.  They can continuously produce
{\it in situ} electrons, distributed throughout the cluster volume 
\cite{Bla99a}.  Secondary electron models can reproduce the
basic properties of the radio halos provided that the strength of the
magnetic field, averaged over the emitting volume, is larger than a few
$\mu$G.  They predict synchrotron power-law spectra which are
independent on cluster location, i.e., do not show any features and/or
radial steepening, and the spectral index values are flatter than
$\alpha$ $\sim$ 1.5 \cite{Bru04a}.  The profiles of the radio
emission should be steeper than those of the X-ray gas
(e.g. \cite{Gov01b}). 
Since the radio emitting electrons originate from protons
accumulated during the cluster formation history, no correlation to
recent mergers is expected, but halos should be present in virtually
all clusters.  Moreover, emission of gamma-rays and of neutrinos is
predicted.

Present observational results, i.e., the behaviour of radio spectra
(see \S~\ref{s:halos}), the association between radio halos and
cluster mergers (\S~\ref{s:connect}), and the fact that halos are
not common in galaxy clusters \cite{Kuo04}, are in favour of
electron reacceleration models.  A two-phase scenario including the
first phase of particle injection, followed by a second phase during
which the aged electrons are reaccelerated by recent merging processes
was successfully applied by Brunetti et al. \cite{Bru01} to the radio halo
Coma C, reproducing its observational properties.

\medskip
\noindent
{\bf Reacceleration}.  In the framework of primary electron models, a
cluster merger plays a crucial role in the energetics of radio
halos. Energy can be transferred from the ICM thermal component to the
non-thermal component through two possible basic mechanisms: 1)
acceleration at shock waves \cite{Sar99, Kes04}; 2)
resonant or non-resonant interaction of electrons with
magneto-hydrodynamic (MHD) turbulence
\cite{Bru01, Bru04b, Pet01, Fuj03}.

Shock acceleration is a first-order Fermi process of great importance
in radio astronomy, since it is recognized as the mechanism responsible
for particle acceleration in the supernova remnants.  The acceleration
occurs diffusively, in that particles scatter back and forth across
the shock, gaining at each crossing and recrossing an amount of energy
proportional to the energy itself.  The acceleration efficiency is
mostly determined by the shock Mach number.  
In the case of radio halos, however, the following arguments
do not favour a connection to  merger shocks:
i) the shocks detected so far with Chandra at the center of several
clusters (e.g. A2744, \cite{Kem04}; A665, 
\cite{Mar01}; 1E0657-56, \cite{Mar02}) have inferred Mach
numbers in the range of $\sim$ 1 - 2.5, which seem too low to
accelerate the radio halo electrons \cite{Gab03}; ii) the
radio emission of halos can be very extended up to large scales, thus
it is hardly associable with localized shocks; iii) the comparison
between radio data and high resolution Chandra X-ray data, performed
by Govoni et al. \cite{Gov04}, shows that some clusters exhibit a spatial
correlation between the radio halo emission and the hot gas regions.
This is not a general feature, however, and in some cases the hottest
gas regions do not exhibit radio emission; iv) the radio spectral
index distribution in A665 \cite{Fer04} shows no evidence of
spectral flattening at the location of the hot shock detected by
Chandra \cite{Mar01}.

Although it cannot be excluded that shock acceleration may be
efficient in some particular regions of a halo (e.g. in A520,
\cite{Mar05}), current observations globally
favour the scenario that cluster turbulence might be the major
mechanism responsible for the supply of energy to the electrons
radiating in radio halos.  Numerical simulations indicate that mergers
can generate strong fluid turbulence on scales of 0.1 - 1 Mpc.
Turbulence acceleration is similar to a second-order Fermi process and is
therefore rather inefficient compared with shock acceleration. The
time during which the process is effective is only a few 10$^8$ 
years, so that the emission is expected to correlate with the
most recent or ongoing merger event.  The mechanism involves the
following steps \cite{Bla04, Bru04b}: 

\noindent
1) the fluid turbulence which is injected into the ICM must 
be converted to MHD turbulence; the
mechanism for this process is not fully established (although the
Lighthill mechanism is mostly used in the recent literature); 

\noindent
2) several types of MHD turbulence modes can be activated (Alfv\`en
waves, slow and fast magnetosonic modes, etc.) and each of them has a
different channel of wave-particle interaction; 

\noindent
3) the cascade process
due to wave-wave interaction, i.e., the decay of the MHD scale size to 
smaller values, must be efficient to produce the MHD scale relevant for
the wave-particle interaction, i.e., for the particle reacceleration
process; 

\noindent
4) the MHD waves are damped because of wave-particle
interaction, so  the reacceleration process could  be eventually reduced.

The particle reacceleration through Alfv\`en waves has the following
limitations: i) the scale relevant for wave-particle interaction is
$\sim$ 1 pc, thus the reacceleration process is efficient only after a
significant cascade process; ii) Alfv\`en waves are strongly damped
through interaction with protons. It follows that if protons are too
abundant in the ICM, they suppress the MHD turbulence and consequently
the reacceleration of electrons.  Brunetti et al. \cite{Bru04b} derived that
the energy in relativistic protons should be $<$ 5-10\% than the
cluster thermal energy to generate radio halos.  In the case of fast
magnetosonic (MS) waves, 
the difficulty of wave cascade to small scales is alleviated
by the fact that their scale of interaction with particles is of the
order of a few kpc. Moreover, the MS wave damping is due to thermal
electrons, and thus hadrons do not significantly affect the electron
reacceleration process \cite{Cas05}.  Therefore, fast MS
waves represent a promising channel for the MHD turbulence
reacceleration of particles.

The emerging scenario is that turbulence reacceleration is the likely
mechanism to supply energy to the radio halos.  All the different
aspects discussed above need to be further investigated in
time-dependent regimes, considering all types of charged particles
\cite{Bru05}, and the contribution of different
mechanisms.

\subsection{Relativistic electrons in radio relics}
\label{s:origre}

Peripheral cluster regions do not host a sufficiently dense thermal
proton population which is required as the target for the efficient 
production of
secondary electrons, and therefore secondary electron models cannot
operate in the case of relics.  
There is increasing evidence that the radio emitting
particles in relics are powered by the energy dissipated in shock
waves produced in the ICM by the flows of cosmological large-scale
structure formation.  The production of outgoing shock waves at the
cluster periphery is indeed observed in numerical simulations of
cluster merger events \cite{Ryu03}.  Because of the electron
short radiative lifetimes, radio emission is produced close to the
location of the shock waves.  This is consistent with the almost 
perpendicular to the merger
axis elongated structure of relics. The
electron acceleration required to produce the relic emission could
result from Fermi-I diffusive shock acceleration of thermal ICM
electrons \cite{En98}, or by adiabatic energization of
relativistic electrons confined in fossil radio plasma, released by a
former active radio galaxy \cite{En01, En02, Hoe04}.  These models
predict that the magnetic field within the relic is aligned with the
shock front, and that the radio spectrum is flatter at the shock edge,
where the radio brightness is expected to decline sharply.

The detection of shocks in the cluster outskirts is presently very
difficult because of the very low X-ray brightness of these
regions. The X-ray data for radio relics are indeed very scarce. The
Chandra data of A754 \cite{Mar03b} indicate that the
easternmost  boundary of the relic coincides with
a region of hotter gas. From XMM  data of the same cluster,
Henry et al. \cite{Hen04} show that the diffuse radio sources
(halo + relic) appear to be associated with high pressure regions.

\subsection{Relativistic electrons in mini-halos}
\label{s:origmh}

Current models for mini-halos involve primary or secondary electrons,
similar to halos.  Gitti et al. \cite{Git02} suggest that the relativistic
primary electrons are continuously undergoing reacceleration due to
the MHD turbulence associated with the cooling flow region.  Pfrommer
\& En{\ss}lin \cite{Pfr04}, on the other hand, discuss the possibility that
relativistic electrons in mini-halos are of secondary origin and thus are
produced by the interaction of cosmic ray protons with the ambient
thermal protons.  Predictions of these models are similar to those of
the halo models. The electron reacceleration model is favoured by the
spectral behaviour of the Perseus mini-halo, i.e. high frequency
steepening and radial spectral steepening \cite{Sij93}, and by the
observed correlation between the mini-halo radio power and the cooling
flow power \cite{Git02}.  Data on this class of
diffuse radio sources, however,  are too poor to draw conclusions.

\section {Cluster magnetic fields}
\label{s:bfield}

The presence of magnetic fields in clusters is directly demonstrated
by the existence of large-scale diffuse synchrotron sources, which
have been discussed in \S~\ref{s:diff}.  In this section, we
present an independent way of obtaining indirect information about the
cluster magnetic field strength and geometry, using data at radio
wavelengths.  This is the analysis of the Faraday rotation of radio
sources in the background of clusters or in the galaxy clusters
themselves.

Measurements of the ICM magnetic fields can also be obtained through
X-ray data from the studies of cold fronts (e.g. \cite{Vik01}) 
and from the detection of non-thermal X-ray emission of
inverse Compton origin, due to scattering of the cosmic microwave
background photons by the synchrotron electrons. The latter emission
can be detected in the hard X-ray domain (e.g. \cite{Fus03}), 
where the cluster thermal emission
becomes negligible.  The studies in the radio band are, however, the
most relevant and provide the most detailed field estimates.

\subsection{Rotation measure}
\label{s:rotmre}

The synchrotron radiation from cosmic radio sources is well known to
be linearly polarized.  A linearly polarized wave of wavelength
$\lambda$, traveling from a radio source through a magnetized medium,
experiences a phase shift of the left versus right circularly
polarized components of the wavefront, leading to a rotation
$\Delta\chi$ of the position angle of the polarization, according to
the law: $\Delta\chi$ = RM $\lambda^2$, where RM is the Faraday
rotation measure.  The RM is obtained as:

\begin{equation}
{\rm RM}=\frac{e^3}{2\pi m_e^2c^4}\int\limits_0^L n_e {\bf B} \cdot d{\bf l}.
\label{rmphys}
\end{equation}

\noindent
In practical units, RM is related to the electron density $n_{\rm e}$,
in units of cm$^{-3}$, 
and to the magnetic field along the line of sight $B_{\parallel}$, 
in units of $\mu$G, through the relation:

\begin{equation}
{\rm RM} = 812\int\limits_0^L n_{\rm e}  B_{\parallel} dl 
~~~~~~{\rm rad \; m^{-2}}\;,
\label{rmprac}
\end{equation}

\noindent
where the path length $l$ is in kpc.
By convention, RM is positive (negative) for a magnetic field
directed toward (away from) the observer.

The RM values can be derived from multi-frequency polarimetric
observations of sources within or behind the clusters, by measuring
the position angle of the polarized radiation as a function of
wavelength.  In general, the position angle must be measured at three
or more wavelengths in order to determine RM accurately and remove the
position angle ambiguity : $\chi_{true}=\chi_{obs}\pm n \pi$.  Once the
contribution of our Galaxy is subtracted, the RM should be dominated
by the contribution of the ICM, and therefore it can be combined with
measurements of $n_{e}$ to estimate the cluster magnetic field along
the line of sight.  This approach can be followed analytically only
for simple distributions of $n_{e}$ and $B$.

A recent technique to analyse and interpret the RM data is the RM
Synthesis, developed by Brentjens \& De Bruyn \cite{Bre05}, which uses the
RM transfer function to solve the $n\pi$ ambiguity related to the RM
computation, and allows one to distinguish the emission as a function of
Faraday depth.

%In simple cases, the strength of the magnetic field can be derived
%by RM measurements with  the following formulas:

Below we present some simple cases, where the strength of the 
magnetic field can be derived by RM measurements:

\medskip
\noindent
{\bf Uniform screen}.
In the simplest approximation of an external screen with uniform magnetic
field, no depolarization is produced and the rotation measure
follows directly from eqn. (\ref{rmprac}):

\begin{equation}
{\rm RM} = 812 \; n_e B_{\parallel} L,
\end{equation}

\noindent
where $n_e$ is in cm$^{-3}$, $B_{\parallel}$ is in $\mu$G, and $L$ is
the depth of the screen in kpc.

\medskip
\noindent
{\bf Screen with tangled magnetic field}.
The effect of a Faraday screen with a tangled magnetic field has been
analyzed by Lawler and Dennison \cite{Law82} and by Tribble \cite{Tri91} in the
ideal case that the screen is made of cells of uniform size, with the same
electron density and the same magnetic field strength, but with field
orientation at random angles in each cell. The observed RM along any
given line of sight will be generated by a random walk process, which results
in a gaussian RM distribution with mean and variance given by:

%\medskip
%
%\par
%~~~~~~~~~~~~~~~~~~~~~

\begin{equation}
\langle {\rm RM} \rangle = 0 \;,\;\;\; \sigma_{\rm RM}^{2}= \langle
{\rm RM}^{2} \rangle = 
812^{2} \; \Lambda_{c} \int ( n_{e} B_{\parallel})^{2}dl~,
\label{sigmarndwalk}
\end{equation}

\noindent
where $n_e$ is in cm$^{-3}$, $B$ is in $\mu$G, and $\Lambda_c$ is the
size of each cell in kpc.  A tangled magnetic field also produces
depolarization (see \cite{Tri91}).

\medskip
\noindent
{\bf Screen with tangled magnetic field and radial gas
density distribution}.
The case of a screen with tangled magnetic field can be treated
analytically if a realistic cluster gas density distribution is
considered, given that the cells have uniform size, the same magnetic
field strength and random field orientation.  If the gas density
follows a hydrostatic isothermal beta model \cite{Cav81}, i.e.,

\begin{equation}
n_e(r)=n_0(1+r^2/r_c^2)^{-\frac{3\beta}{2}},
\label{king}
\end{equation}

\noindent
where $n_0$ is the central
electron density, and $r_c$ is the core radius of the gas distribution, 
the value of the RM variance is given by:

\begin{equation}
\sigma_{\rm RM}(r_{\perp})= {{K B n_{0}  r_c^{\frac{1}{2}} \Lambda_{c}^{\frac{1}{2}} }
 \over {(1+r_{\perp}^2/r_c^2)^{\frac{(6\beta -1)}{4}}}} \sqrt 
{{\Gamma(3\beta-0.5)}\over{\Gamma(3\beta)}},
\label{felten}
\end{equation}

\noindent
where $r_{\perp}$ is the projected distance from the cluster centre
and $\Gamma$ indicates the Gamma function.  The constant $K$ depends
on the integration path over the gas density distribution: $K$ = 624,
if the source lies completely beyond the cluster, and $K$ = 441 if the
source is halfway through the cluster.  

\noindent
For $\beta$=0.7 the previous formula becomes:

\begin{equation}
 \sigma_{\rm RM} \approx {{575 B}\over
{(1+r^2/r_c^2)^{0.8}}}  n_0  M^{\frac{1}{2}}l. 
\end{equation}

\noindent
Note that depolarization is also produced, due to the fact that the magnetic
field is tangled.

\subsection{Current results from RM studies}
\label{s:rm}

\begin{figure}[t]
\centering
\includegraphics[height=16pc]{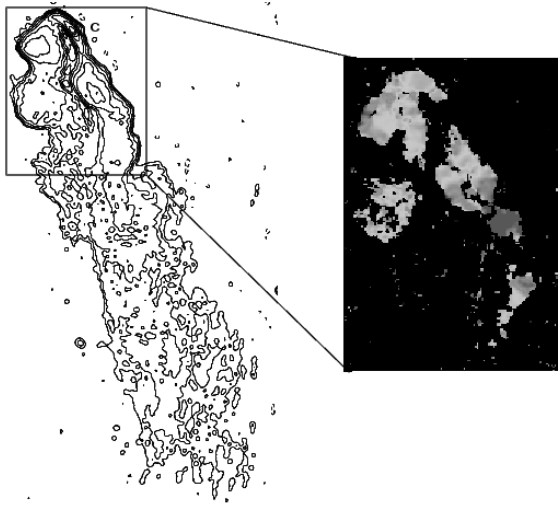}
\caption{VLA contour plot of the tailed radio galaxy
 0053-015 in A119 at 1.4 GHz (left), and RM image (right).
The values of RM range between --350 rad m$^{-2}$ and +450 rad m$^{-2}$,
with $\langle$RM$\rangle$ = + 28 rad m$^{-2}$,
and a dispersion of $\sigma_{RM}$ = 152 rad m$^{-2}$. They 
show fluctuations on scales of $\sim$ 3.5 arcsec
\cite{Fer99}.
}
\label{rmsource}
\end{figure}

Cluster surveys of the Faraday rotation measures of polarized radio
sources both within and behind clusters provide an important probe of
the existence of intracluster magnetic fields.  The RM values derived
in background or embedded cluster sources are of the order of tens to
thousands rad m$^{-2}$ (an example is shown in Fig. \ref{rmsource}).  The
observing strategy to derive information on the magnetic field
intensity and structure is twofold: i) obtain the average value of the
RM of sources located at different impact parameters of the cluster,
ii) derive maps of the RM of extended radio sources, to evaluate the
$\sigma$ of the RM distribution.

\begin{figure}[t]
\centering
\includegraphics[height=16pc]{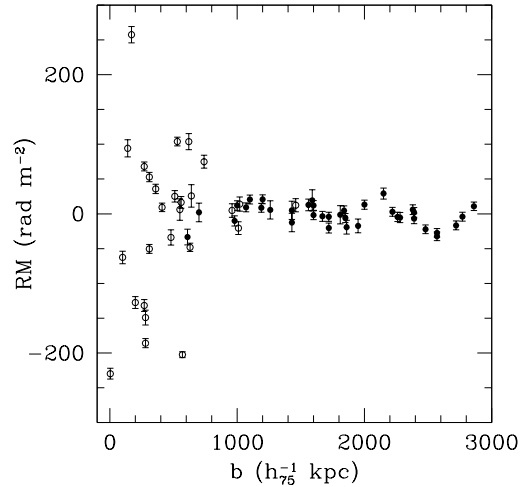}
\caption{
Galaxy-corrected rotation measure plotted as a
function of source impact parameter in kiloparsecs for the sample of
sources from Clarke et al. \cite{Cla01}. Open dots refer to cluster sources,
closed dots to control sources.
}
\label{rmstat}
\end{figure}

Studies have been carried out on both statistical samples and
individual clusters (see e.g. the review by Govoni \& Feretti \cite{Gov04a} and
references therein).  Kim et al. \cite{Kim91} analyzed the RM of 53 radio
sources in and behind clusters and 99 sources in a control sample.
This study, which contains the largest cluster sample to date,
demonstrated that $\mu$G level fields are widespread in the ICM.  In
a more recent statistical study, Clarke et al. \cite{Cla01} analyzed RMs for
a representative sample of 16 cluster sources, plus a control sample,
and found a statistically significant broadening of the RM
distribution in the cluster sample, and a clear increase in the width
of the RM distribution toward smaller impact parameters (see
Fig. \ref{rmstat}).  They derived that the ICM is permeated with a
high filling factor of magnetic fields at levels of 4 - 8 $\mu$G and
with a correlation length of $\sim$15 kpc, up to $\sim$0.75 Mpc from
the cluster centre. The results are confirmed by new data on an
expanded sample \cite{Cla04}.

The first detailed studies of RM within individual clusters have been
performed on cooling core clusters, owing to the extremely high RMs of
the powerful radio galaxies at their centres (e.g., Hydra A,
\cite{Tay93}; 3C295, \cite{All01}).  High values of the magnetic
fields, up to tens of $\mu$G, have been obtained, but they only refer
to the innermost cluster regions.  Studies on larger areas of clusters
have been carried out e.g. for Coma \cite{Fer95}, A119
\cite{Fer99}, A514 \cite{Gov01c}, 3C129 \cite{Tay01}.

Overall, the data are consistent with cluster atmospheres containing
magnetic fields in the range of 1-5 $\mu$G, regardless of the presence
or not of diffuse radio emission. At the centre of cooling core
clusters, magnetic field strengths can be larger by more than a factor
of 2.  The RM distributions are generally patchy, indicating that
large-scale magnetic fields are not regularly ordered on cluster
scales, but have coherence scales between 1 and 10 kpc.  
In most clusters the magnetic fields are not dynamically important, with
magnetic pressures much lower than the thermal pressures, but the
fields may play a fundamental role in the suppression of the particle
thermal conduction \cite{Cha99} and in the energy budget of
the ICM.

\subsection{Magnetic field structure}
\label{s:bstruct}

The simplest model is a uniform field throughout the cluster. However,
this is not realistic: if the field values detected at the cluster
centres extend over several core radii, up to distances of the
order of $\sim$ Mpc, then the magnetic pressure would exceed the thermal
pressure in the outer parts of the clusters.  The magnetic field
intensity is likely to decrease with the distance from the cluster
centre, as derived in Coma \cite{Bru01}.  This is also
predicted as a result of compression of the thermal plasma during the
cluster gravitational collapse, where the magnetic field-lines are
frozen into the plasma, and compression of the plasma results in
compression of flux lines. As a consequence of magnetic flux
conservation, the expected growth of the magnetic field is
proportional to the gas density as B $\propto \rho^{2/3}$.  

Dolag et al. \cite{Dol01} showed that in the framework of hierarchical
cluster formation, the correlation between two observable parameters,
the RM and the X-ray surface brightness, is expected to reflect the
correlation between  the magnetic field and
gas density. Therefore, from the analysis of the RM versus X-ray
brightness it is possible to infer the trend of magnetic field versus
gas density.  The application of this approach has been possible so
far only in A119, giving the radial profile of the magnetic field as
$B \propto n_e^{0.9}$ \cite{Dol01}. The
magnetic field decline with radius is confirmed in this case.

Another important aspect to consider is the structure in the cluster
magnetic field, i.e. the existence of filaments and flux ropes
\cite{Eil99}.  
The magnetic field structure can be investigated by deriving
the power spectrum of the field fluctuations, defined as:
$|B_{\kappa}|^2 \propto {\kappa}^{-n}$, where ${\kappa}$ represents
the wave number of the fluctuation scale.  By using a semi-analytic
technique, En{\ss}lin \& Vogt \cite{En03} and Vogt \& En{\ss}lin \cite{Vog03}
showed that the magnetic field power spectrum can be estimated by
Fourier transforming RM maps, if very detailed RM images are available.
Alternatively, a numerical approach using Monte Carlo simulations has
been developed by Murgia et al. \cite{Mur04} to reproduce the rotation
measure and the depolarization produced by magnetic field with different
power spectra.

\subsection{Reconciling values derived with different 
approaches}
\label{s:reconcil}

The cluster magnetic field values obtained from RM arguments are about
an order of magnitude higher than those derived from both the
synchrotron diffuse radio emission (\S~\ref{s:halos}) and the
inverse Compton (IC) hard X-ray emission (e.g. \cite{Fus03}).
The discrepancy can be alleviated by taking into account that:

\begin{itemize}
\item estimates of equipartition fields rely on
several assumptions (see \S~\ref{s:equip}); 

\item Goldsmith \& Rephaeli \cite{Gol93} 
suggested that the IC estimate is typically expected
to be lower than the Faraday rotation estimate, because of the 
spatial profiles of the magnetic field and gas density. For example,
if the magnetic field strength has a radial decrease, most of the IC
emission will come from the weak field regions in the outer parts of
the cluster, while most of the Faraday rotation and synchrotron
emission occurs in the strong field regions in the inner parts of the
cluster; 

\item it has been shown that IC models which
include the effects of aged electron spectra, combined with
the expected radial profile of the magnetic field, and anisotropies in
the pitch angle distribution of the electrons, allow higher values of
the ICM magnetic field in better agreement with the Faraday rotation
measurements \cite{Bru01, Pet01}; 

\item the magnetic field may show complex structure, as filamentation
and/or substructure with a range of coherence scales (power spectrum).  
Therefore, the
RM data should be interpreted using realistic models of the cluster
magnetic fields (see \S~\ref{s:bstruct});

\item Beck et al. \cite{Bec03} pointed out that field estimates derived from RM
may be too large in the case of a turbulent medium where small-scale
fluctuations in the magnetic field and the electron density are highly
correlated ; 

\item it has been recently pointed out that in some cases a
radio source could compress the gas and fields in the ICM to produce
local RM enhancements, thus leading to overestimates of the derived
ICM magnetic field strength \cite{Rud03}; 

\item evidence
suggests that the magnetic field strength will vary depending on the
dynamical history and location within the cluster. A striking example
of the variation of magnetic field strength estimates for various
methods and in various locations throughout the cluster is given in
\cite{Joh04}.
\end{itemize}

Future studies are needed to shed light on these issues and improve
our current knowledge on the strength and structure of the magnetic fields. 

\subsection {Origin of cluster magnetic fields}
\label{s:origb}

The field strengths that we observe in clusters greatly exceed the
amplitude of the seed fields produced in the early universe, or fields
injected by some mechanism by high redshift objects.
There are two basic possibilities for their origin: 

\noindent
1) ejection from
galactic winds of normal galaxies or from active and starburst
galaxies \cite{Kro99, Vol99};

\noindent
2) amplification of seed fields during the cluster formation
process.

Support for a galactic injection in the ICM comes from the evidence
that a large fraction of the ICM is of galactic origin, since it
contains a significant concentration of metals.  However, fields in
clusters have strengths and coherence size comparable to, and in some
cases larger than, galactic fields \cite{Gra01}.
Therefore, it seems quite difficult that the magnetic fields in
the ICM derive purely from ejection of the galactic fields,
without invoking other amplification mechanisms \cite{DeY92, Reph88}.  

Magnetic field amplification is likely to occur during the cluster
collapse, simply by compression of an intergalactic field. Clusters
have present day overdensities $\rho \sim 10^3$ and in order to get
$B_{\rm ICM} > 10^{-6}$ G by adiabatic compression ($B \propto
\rho^{2/3}$) requires intergalactic (seed) fields of at least
10$^{-8}$ G. These are somewhat higher than current limits derived in
the literature \cite{Bar97, Bla99b}.  A possibile way
to obtain a larger field amplification is through cluster mergers. Mergers
generate shocks, bulk flows and turbulence within the ICM.  The first
two of these processes can result in some field amplification simply
through compression. However, it is the turbulence which is the most
promising source of non-linear amplification.  MHD calculations have
been performed \cite{Dol99, Roe99, Sub05} to investigate the evolution of magnetic fields.
The results of these simulations show that cluster mergers can
dramatically alter the local strength and structure of cluster-wide
magnetic fields, with a strong amplification of the magnetic field
intensity.  Shear flows are extremely important for the amplification
of the magnetic field, while the compression of the gas is of minor
importance.  The initial field distribution at the beginning of the
simulations at high redshift is irrelevant for the final structure of
the magnetic field. The final structure is dominated only by the
cluster collapse.  Fields can be amplified from initial values of
$\sim$ 10$^{-9}$ G at $z=15$ to $\sim$ 10$^{-6}$ G at the present epoch
\cite{Dol99}.  Roettiger et al. \cite{Roe99} found a significant evolution of
the structure and strength of the magnetic fields during two distinct
epochs of the merger evolution. In the first, the field becomes quite
filamentary as a result of stretching and compression caused by shocks
and bulk flows during infall, but only minimal amplification
occurs. In the second, amplification of the field occurs more rapidly,
particularly in localized regions, as the bulk flow is replaced by
turbulent motions.  Mergers change the local magnetic field strength
drastically, but also the structure of the cluster-wide field is
influenced. At early stages of the merger the filamentary structures
prevail. This structure breaks down later ($\sim$ 2--3 Gyr) and leaves
a stochastically ordered magnetic field.  Subramanian et al \cite{Sub05}
argue that the dynamo action of turbulent motions in the intracluster
gas can amplify a random magnetic field by a net factor of 10$^4$ in 5
Gyr. The field is amplified by random shear, and has an intermittent
spatial distribution, %making possibly filaments.
possibly producing filaments.

\section{Radio emission from cluster radio galaxies}
\label{s:rg}

Recent results on the thermal gas in clusters of galaxies has revealed
a significant amount of spatial and temperature structure, indicating
that clusters are dynamically evolving by accreting gas and galaxies
and by merging with other clusters/groups (roughly every few Gyrs).
Simulations suggest that the ICM within clusters is violent, filled
with shocks, high winds and turbulence.  This gas can interact with a
radio source in different ways: modifying its morphology via ram
pressure, confining the radio lobes, possibly feeding the active
nucleus.  We discuss below some of the recent results on these topics
(see also the review of Feretti \& Venturi \cite{Fer02b}).

\subsection{Interaction between the radio galaxies and the ICM}
\label{s:natwat}

\begin{figure}[t]
\centering
\includegraphics[width=4.5cm]{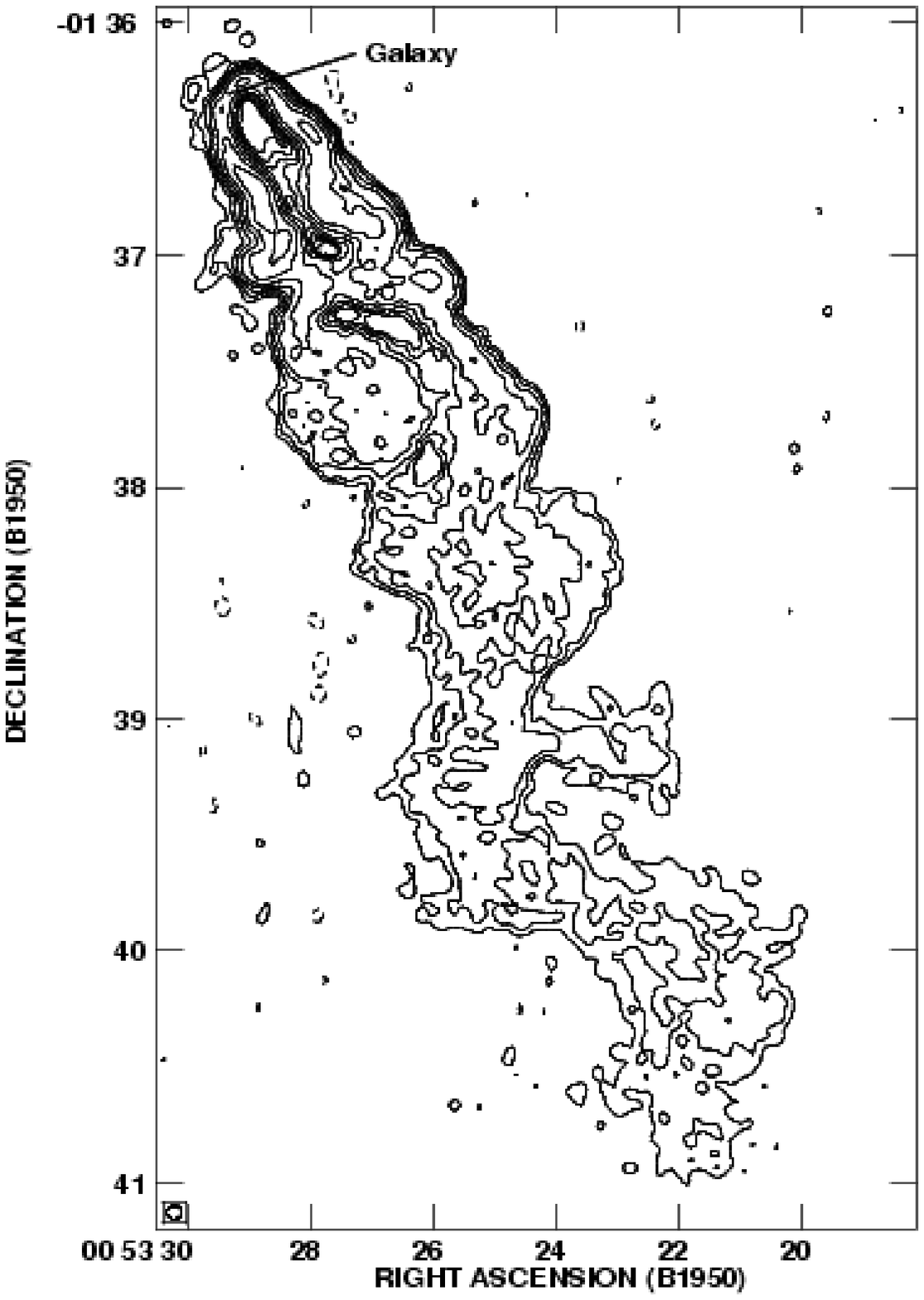}
\includegraphics[width=6.5cm]{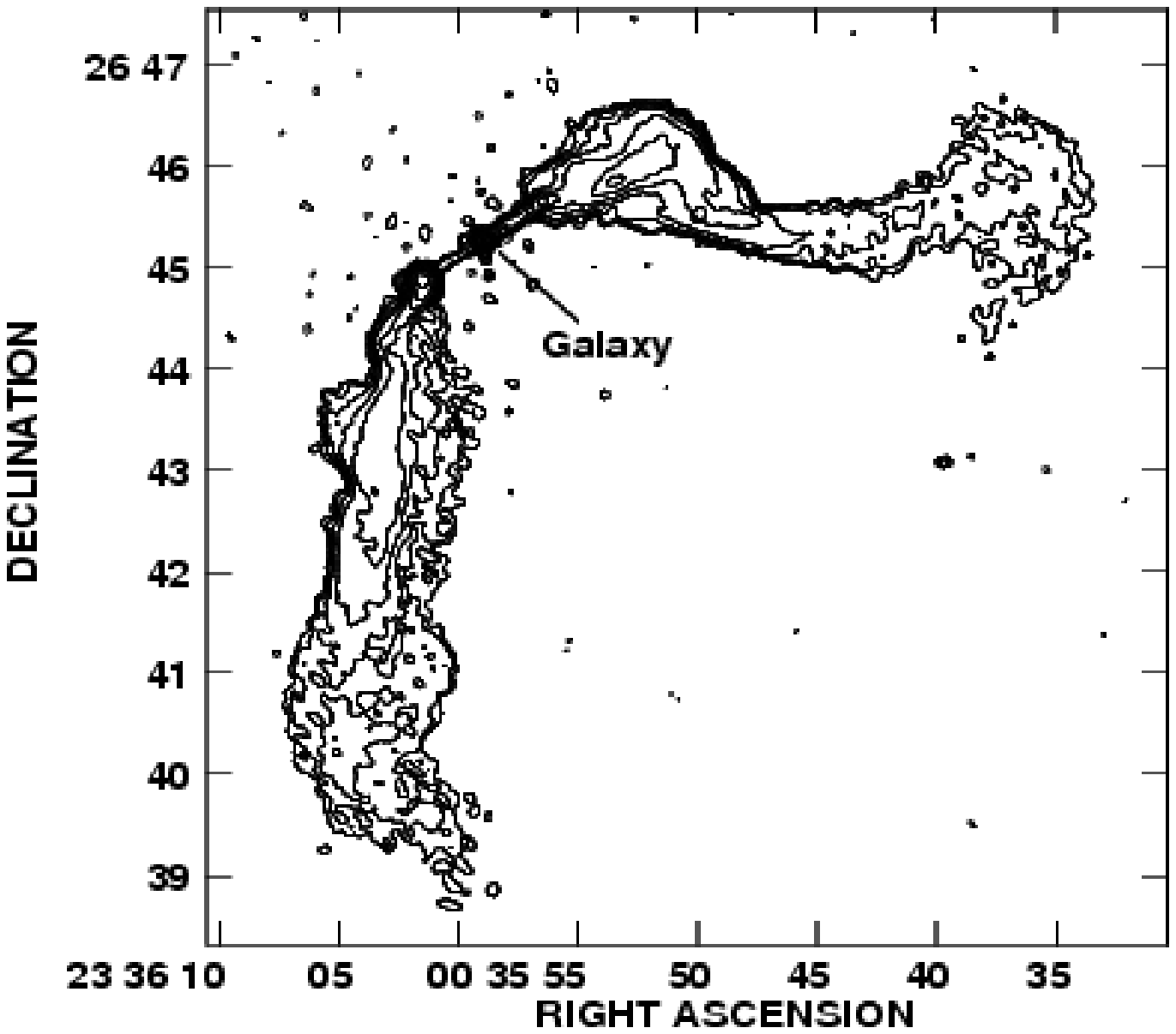}
\caption{
Examples of tailed radio galaxies: 
the NAT  0053-016 in the cluster A119 ({\bf left panel}) and 
the WAT 3C465 in the cluster  A2634 ({\bf right panel}).
The location of the optical galaxy is indicated.}
\label{code}
\end{figure}

{\bf Tailed radio galaxies.}  A dramatic example of the interaction of
the radio galaxies with the ICM is represented by the tailed radio
galaxies, i.e. low-power radio sources (FR~I type, \cite{Fan74}) 
where the large scale low-brightness emission is bent towards the
same direction, forming features similar to tails.  These radio
galaxies were originally distinguished in two classes: narrow-angle
tailed sources (NAT), which are "U" shaped with a small angle between
the tails, and wide-angle tailed sources (WAT), which are "V" shaped
with a larger angle between the tails (see Fig. \ref{code}).  We note
that distortions in powerful radio galaxies (FR~II type,
\cite{Fan74}) are marginal and only present in weak structures.

The standard interpretation of the tailed radio morphology is that the
jets are curved by ram pressure from the high-velocity host galaxy
moving through the dense ICM, whereas the low brightness tails are
material left behind by the galaxy motion.  The ram-pressure model was
first developed by Begelman et al. \cite{Beg79}.  Following dynamical
arguments, the bending is described by the Euler equation:

\begin{equation}
R \sim h \left(\frac{\rho_j}{\rho_e}\right) \left(\frac{v_j}{v_g}\right)^2 ,
\end{equation}

\noindent
where R is the radius of curvature, $\rho$ is density, $v$ is velocity
(the subscript $j$ refers to the jet, $e$ to the external medium, $g$
to the galaxy) and $h$ is the scale height over which the ram pressure
is transmitted to the jets.  Thus, from the jet bending, important
constraints on both the jet dynamics and the ICM can be placed.  In
some cases there is evidence that the radio jets travel first through
the galactic atmosphere and then are sharply bent at the transition
between the galactic atmosphere and the ICM.  Bends can occur very
close to the nucleus, as in NGC 4869 in the Coma cluster \cite{Fer90}, 
indicating that the bulk of interstellar medium has been
stripped by the galaxy during its motion.

\begin{figure}[t]
\centering
\includegraphics[height=9.0cm]{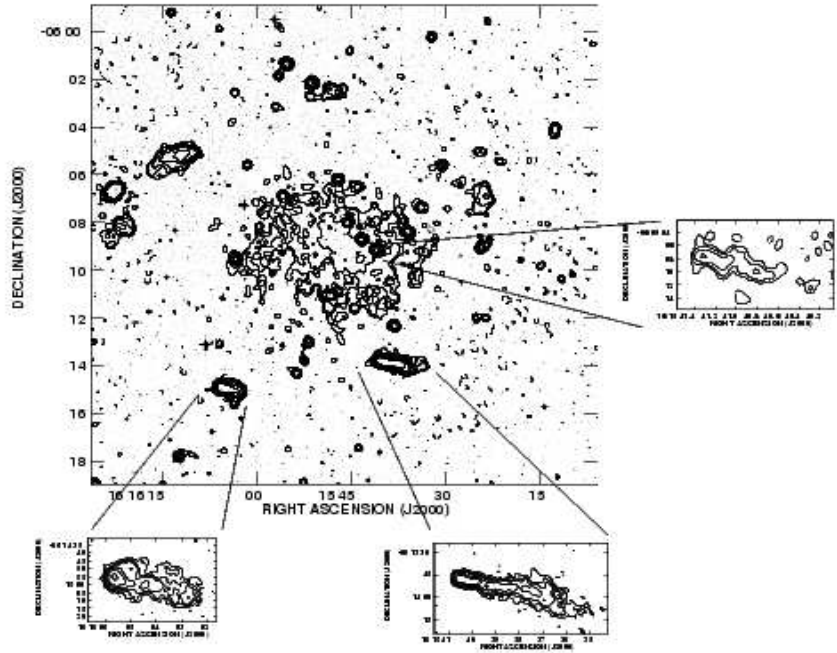}
\caption{
Radio image of the cluster A2163 at 1.4 GHz, with angular 
resolution of 15\arcsec \cite{Fer01}.
The structure of tailed radio galaxies as detected at higher resolution
is shown in the insets. The tails are all oriented in the same
direction. }
\label{nat2163}
\end{figure}

In general, the ram-pressure model can explain the radio jet
deflection when the galaxy velocity with respect to the ICM is of the
order of $\sim$ 1000 km s$^{-1}$. Therefore, it can successfully
explain the structure of NAT sources, which are indeed identified with
cluster galaxies located at any distance from the cluster centre and
thus characterized by significant motion.
However, Bliton et al. \cite{Bli98} derived that NATs are preferentially
found in clusters with X-ray substructure. Additionally, NAT galaxies
tend to have, on average, velocities similar to those of typical
cluster members, instead of high peculiar motions expected if NATs
were bent only from ram pressure. Thus, they suggested a new model for the
NAT formation, in which NATs are associated with dynamically complex
clusters with possible recent or ongoing cluster-subcluster mergers.
The U-shaped morphology is then suggested to be produced, at least in
part, by the merger-induced bulk motion of the ICM bending the jets.
This is supported, in some clusters, by the existence of NAT radio
galaxies with their tails oriented in the same direction (e.g., A2163,
Fig. \ref{nat2163}; A119, \cite{Fer99}), since 
it seems unlikely that their parent galaxies are all moving
 towards the same direction.

\begin{figure}[t]
\centering
\includegraphics[height=5.0cm]{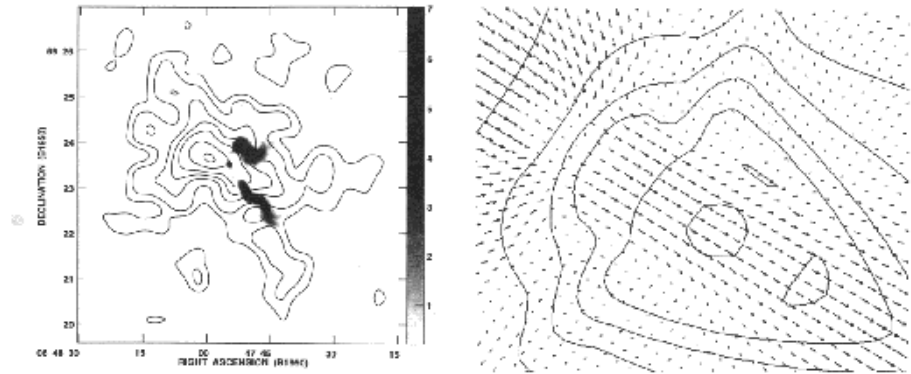}
\caption{{\bf Left panel}: Overlay of the grey-scale radio image
of the WAT source in A562 onto the ROSAT X-ray surface 
brightness contour image of the same cluster. 
{\bf Right panel}: Overlay of a synthetic X-ray image of a cluster merger
onto a velocity vector field that represents the gas velocity. Note that
the X-ray contours in the left panel look very similar to the synthetic
X-ray image and that the radio tails are in the direction of 
the gas velocity (from \cite{Gom97}).}
\label{watmo}
\end{figure}

The interpretation of WAT sources may be problematic in the framework
of the ram-pressure model, since these sources are generally
associated with dominant cluster galaxies moving very slowly (\ltsim ~100
km s$^{-1}$) relative to the cluster velocity centroid.  Such slow
motion is insufficient to bend the jets/tails of WATs to their
observed curvature by ram pressure.  It has therefore been suggested that
WATs must be shaped mostly by other ram-pressure gradients
not arising from the motion of the host galaxy, but produced by mergers
between clusters \cite{Lok95, Gom97}.  Numerical
simulations lead support to this idea: peak gas velocities are found well in
excess of 1000 km s$^{-1}$ at various stages of the cluster merger
evolution, which generally do not decay below 1000 km
s$^{-1}$ for nearly 2 Gyr after the core passage. This is consistent
with the observations, as modelled in the cluster A562
(Fig. \ref{watmo}).

\medskip
\noindent
{\bf Radio emission in X-ray cavities}.
A clear example of the interaction between the radio plasma and the
hot intracluster medium was found in the ROSAT image of the Perseus
cluster \cite{Boh93}, where X-ray cavities associated
with the inner radio lobes to the north and south of the bright
central radio galaxy 3C84 have been first detected.  The high spatial
resolution of the Chandra X-ray Observatory has confirmed the presence
of such X-ray holes \cite{Fab00}, coinciding with the radio
lobes and showing rims cooler than the surrounding gas.  Chandra has
permitted the detection of X-ray deficient bubbles in the inner region
of many cooling flow clusters, e.g.,  Hydra A, A2052, A496, A2199,
RBS797. These features are discussed by C. Jones et al. in 
this volume.

\subsection{Trigger of radio emission}
\label{s:trigger}

An important issue is to understand whether and how the cluster
environment plays any role in the statistical radio properties of
galaxies, in particular their probability of forming radio sources.
The high density of galaxies within clusters, especially in the
innermost cluster regions, and the peculiar velocities of galaxies,
most extreme in merging clusters, enhance the probability of
galaxy-galaxy interactions. These special conditions raise the
questions whether cluster galaxies have enhanced probability of
developing a radio source, and whether they tend to have more powerful
and long lived radio emission. 

A powerful statistical tool to address the above questions is the
radio luminosity function (hereinafter RLF).  The fractional RLF is
defined as:

\begin{equation}
f_i(P,z) = \frac{\rho_i(P,z)}{\phi_i(z)},
\end{equation}

\noindent
where $\phi_i(z)$ is the density of objects of a particular class $i$
at the epoch $z$, and $\rho_i(P,z)$ is the density of the same class
objects showing a radio emission of power P.  The fractional RLF,
$f(P)$, thus represents the probability that a galaxy in a defined
sample at a given epoch emits with radio power in the interval $P \pm
dP$. From an operational point of view, the RLF can be expressed as:

\begin{equation}
f(P) = \frac {n(\Delta P_i)} {N(\Delta P_i)},
\end{equation}

\noindent
where $n$ and $N$ are respectively the number of detected radio galaxies
in the power interval $\Delta P_i$
and the total number of optical galaxies which could have
been detected in the same power bin. 
The integral form of the RLF $F(>P$) can be obtained
simply summing over all radio power intervals up to the power $P$.
In order to take into account the correlation between the optical and radio
properties of galaxies, it is useful to introduce the bivariate
luminosity function $f(P,M)$, which  gives the probability that 
a galaxy with absolute magnitude in the range $M \pm dM$ is radio emitting 
in the radio power range $P \pm dP$.

The RLF of galaxies in clusters has been first investigated by Fanti
\cite{Fanti84}, and latter by Ledlow and Owen \cite{Led96}.  The most striking
result is that statistical properties of radio galaxies are
surprisingly similar for sources both inside and outside rich
clusters.  For both cluster and non-cluster galaxies, the only 
parameter relevant for the radio emission
seems to be the optical magnitude, i.e. brighter galaxies
have a higher probability of developing a radio galaxy.  Furthermore,
the radio luminosity function is independent on richness class,
Bautz-Morgan or Rood-Sastry cluster class.
Recently, Best et al. \cite{Bes05} demonstrated that, while
the radio power of a radio galaxy does not correlate to its mass, 
the probability  of a galaxy to become a radio source is a very 
strong function of both stellar mass and central black hole mass.

It is still under debate whether the universality of the local RLF for
early type galaxies can be applied also to merging clusters.
According to some authors (e.g. \cite{Ven01, Gia04}) 
the enhanced probability of galaxy interaction in merging
clusters has no effect on the probability of galaxies to develop a
radio active galactic nucleus in their centres.

In the cluster A2255, instead, Miller \& Owen \cite{Mil03} found an excess
of powerful radio galaxies, which is interpreted as due to the
dynamical state of the cluster.  Best \cite{Bes04} showed that the
fraction of radio loud AGN appears to be strongly dependent upon the
large scale environment of a galaxy. 
 This supports the argument that a merger process may affect the AGN
activity, since infalling galaxies or galaxy groups more likely
produce galaxy interactions or galaxy-galaxy  mergers which can
trigger the AGN activity. The
effect of cluster merger processes on the trigger of radio emission
would imply an enhanced number of radio source in cluster at high
redshift, i.e. at the earlier epochs when the clusters are being
assembled.  These issues are under investigation. The result of
Branchesi et al. \cite{Bra05} points to a higher number of radio galaxies
in distant clusters, although with poor statistics.  In conclusion,
whereas the ICM in clusters has strong effect on the structures of
radio galaxies, the probability of forming radio sources is likely
unaffected by the cluster environment, but may be affected by cluster
mergers.

Other effects of the interaction between galaxies and ICM, as
the trigger of star formation, the gas stripping, HI deficiency, etc.,
are discussed by other authors in this volume.

\section*{Acknowledgements}

LF is grateful to the organizers David Hughes, Omar L\'opez-Cruz and
Manolis Plionis for their invitation to this stimulating and very
interesting school.  We acknowledge Gianfranco Brunetti for
illuminating discussions on the models of relativistic particle origin and
re-acceleration.

% BibTeX users please use
% \bibliographystyle{}
% \bibliography{}

\end{document}